\documentclass[preprint]{aastex}
\usepackage{epsfig}

\shorttitle{The radio surface brightness to diameter relation for galactic supernova remnants: sample selection and robust analysis with
various fitting offsets}
\shortauthors{Pavlovi\' c et al.}

\begin{document}


\title{THE RADIO SURFACE BRIGHTNESS TO DIAMETER RELATION FOR GALACTIC SUPERNOVA REMNANTS: SAMPLE SELECTION AND ROBUST ANALYSIS WITH
VARIOUS FITTING OFFSETS}


\author{M. Pavlovi\'{c}$^{1}$, D. Uro{\v s}evi{\'c}$^{1,2}$, B. Vukoti{\'c}$^{3}$, B. Arbutina$^{1}$, \"{U}. D. G\"{o}ker$^{4}$,} 

\affil{$^{1}$Department of Astronomy, Faculty of Mathematics,
University of Belgrade, Studentski trg 16, 11000 Belgrade, Serbia}
\email{marko@math.rs}

\affil{$^{2}$Isaac Newton Institute of Chile, Yugoslavia Branch}

\affil{$^{3}$Astronomical Observatory, Volgina 7, 11060 Belgrade 38, Serbia}

\affil{$^{4}$Physics Department, Bo\u{g}azi\c{c}i University, Bebek, 34342, Istanbul, Turkey}




\begin{abstract}
In this paper we present new empirical radio surface brightness-to-diameter ($\Sigma-D$) relations for supernova
remnants (SNRs) in our Galaxy. We also present new theoretical derivations of the $\Sigma-D$ relation based on equipartition or on constant ratio between cosmic rays and magnetic field energy.
A new calibration sample of 60 Galactic SNRs with independently determined distances is created.
Instead of (standard) vertical regression, used in previous papers, different fitting procedures are applied to
the calibration sample in the $\log \Sigma - \log D$ plane.
Non-standard regressions are used to satisfy the requirement that values of
parameters obtained from the fitting of $\Sigma-D$ and $D-\Sigma$ relations
should be invariant within estimated uncertainties.
We impose symmetry between $\Sigma-D$ and $D-\Sigma$ due to the existence of large scatter in both $D$ and $\Sigma$.
Using four fitting methods which treat $\Sigma$ and $D$ symmetrically, different $\Sigma-D$ slopes $\beta$ are obtained for the calibration sample.
Monte Carlo simulations verify that the slopes of the empirical $\Sigma-D$ relation should
be determined by using orthogonal regression, because of its good performance for data sets with severe scatter.
The slope derived here ($\beta=4.8$) is significantly
steeper than those derived in previous studies.
This new slope is closer to the updated theoretically predicted surface brightness-diameter slope in the radio range for the Sedov phase. We also analyze the empirical $\Sigma-D$ relations for SNRs in the dense environment of molecular clouds and for SNRs
evolving in lower-density interstellar medium.
Applying the new empirical relation to estimate distances of Galactic SNRs results in a dramatically changed distance scale.
\end{abstract}





\section{Introduction}

Supernova remnants (SNRs) are the main sources of kinetic energy and heat for
the interstellar medium (ISM). They also contribute to the acceleration of cosmic rays.
The total number of Galactic
SNRs is predicted to be between 1000 and 10,000 \citep{berk84,lietal91,tammetal94}.
Assuming a typical evolution timescale ($\sim 10^{5}$ yr)
of SNRs before they merge with the ISM and an event rate of two supernovae (SNe) per century
in the Milky Way \citep{drag99}, $\sim$ 2000 SNRs are expected in our Galaxy.
However, just 274 SNRs have been identified in our galaxy from their emission line spectra,
radio and X-ray radiation \citep{green09}. Such a great deficit points to the
selection effects in the past surveys. All known SNRs are sources of radio-synchrotron emission.

SNRs were often discovered in radio surveys of the Galactic plane. Because of the surface-brightness
limit of previous surveys, more faint or confused SNRs await discovery. The dominant selection effects are those that are applicable at radio wavelengths. To put it simple, three selection effects apply to the identification of Galactic SNRs \citep[e.g.][]{green91}: (i) difficulty in identifying faint remnants,
(ii) difficulty in identifying small angular size remnants and (iii) the absence of uniform coverage of the sky. Additionally, Malmquist bias\footnote{The volume selection effect --- brighter objects are favored in flux density limited surveys.} acts severely in the Galactic samples making them incomplete. In the case of extragalactic SNRs, the selection effects are different \citep[see][]{ur05}.

The important and very difficult task of determining distances to Galactic SNRs is often based on
observations in radio domain. It is possible to determine direct distances from
historical records of supernovae, proper motions and radial
velocities, kinematic observations, coincidences with H I, H II and molecular clouds, OB
associations and pulsars \citep{green84}. Where direct distance determination is not possible, estimates can be made for shell SNRs
using the radio surface brightness-diameter relationship ($\Sigma-D$). The mean surface brightness at
a specific radio frequency, $\Sigma_{\nu}$, is a distance-independent parameter and is an important characteristic of an SNR \citep{skl60a}.
The radio surface brightness at frequency $\nu$ to diameter relation for SNRs has been written in general form as:
\begin{equation}
\Sigma_{ \nu }(D) = AD^{-\beta},
\label{Sigma-D}
\end{equation}
\noindent where $A$ depends on the properties of the SN explosion and interstellar medium, such as the SN energy
of explosion, the mass of the ejected matter, the density of the ISM, the magnetic field strength, etc., while
$\beta$ is thought to be independent of these properties \citep{ariur05}. For empirical relations, parameters $A$ and $\beta$ are obtained by fitting the data for a sample of SNRs of known distances.

According to the existing convention, SNRs are classified into three basic types: shell remnants,
plerionic or filled-center remnants, and composite remnants. \citet{alla1983a, alla1983b, alla1986a, alla1986b} showed that $\Sigma-D$ relation is only applicable to shell-type SNRs. However
it can also be
used for composite remnants if we separate the flux from the surrounding shell and the flux from the central regions.

The first theoretical $\Sigma-D$ relation was derived by \citet{skl60a}. He was also the first to propose the use of
this relation as a method for determining SNR distances \citep{skl60b}. An updated theoretical derivation of this
relation for shell-like SNRs was done by \cite{bif04}. The first empirical relations were derived by
\citet{piw68}, \citet{mil70} and later by \citet{cic76}, \citet{mil79} and many other authors. An important Galactic
relation, frequently used in the last decade, was derived by Case \& Bhattacharya (1998, hereafter CB98). The updated Galactic $\Sigma-D$ relation
were derived by \citet{gus03} and \citet{xu05}. More than five decades after the first published
study of the relation, it continues to evolve both in theoretical and empirical aspects. A detailed review of Galactic and
extragalactic empirical $\Sigma-D$ relations was presented by \citet{ur2002}.

Although many authors have improved the theoretical and empirical $\Sigma-D$ relations, there are still some doubts when this relation
is applied to determine distances to individual remnants. \citet{green84} presented a critical analysis, noting
that remnants with reliable distances have widespread intrinsic properties leading to significant dispersion in the $\Sigma-D$ diagram. One of the
main problems is the high level of uncertainty for independently determined distances of SNRs, which are used for $\Sigma-D$ calibration. Also, a variety of selection effects may be present in the data samples for both Galactic and extragalactic SNRs. \citet{arbo04} concluded that
evolutionary paths may differ from remnant to remnant because of the potentially wide range of intrinsic properties of
supernova (SN) explosions (and progenitor stars) and the interstellar medium (ISM) into which they expand. \citet{ariur05} showed that
different $\Sigma-D$ relations can be constructed for two different samples, remnants in a dense environment of molecular
clouds and SNRs evolving in ISM of lower density.

A similar relation can be also used for composite SNRs, but only few of those remnants have known distances, and it is
difficult to obtain a $\Sigma-D$ relation for them. The $\Sigma-D$ relation is often the only method for distance determination for shell SNRs. Despite inherent uncertainties and assumptions of this method, we suggest that the often-used $\Sigma-D$ relation by CB98
should be updated because of the significant increase in the number of SNR calibrators. Additionally, different fitting procedures have to be used for establishing proper $\Sigma-D$ calibration -- we emphasize that in all previous papers the standard fitting procedure, based on vertical (parallel to $y$-axes) $\chi^{2}$ regression, was being used to
derive empirical $\Sigma-D$ relations for the Galactic SNR samples (e.g. \citealt{piw68}; \citealt{cic76}; \citealt{mil79}; CB98; \citealt{gus03}; \citealt{xu05}).
In this paper, we select calibrators from Green's SNR catalog \citep{green09}
based on literature up to the end of 2008.\footnote{A Catalogue of Galactic Supernova Remnants (2009 March version),
Astrophysics Group, Cavendish Laboratory, Cambridge, United Kingdom (also available on the World Wide Web at http://www.mrao.cam.ac.uk/surveys/snrs/).}

The $\Sigma-D$ relation has two particular uses: first, the derivation of diameters (and hence distances)
and second, parametrization of the relationship between surface
brightness and diameter, for comparison with models and theories. For the former, the measured
value of $\Sigma$ is used to predict $D$, and for this case
a least squares fit minimizing the deviations in
$\log{D}$ should be used.
However, \citet{ur2010} suggested the use of orthogonal
regression fitting procedure for obtaining the empirical $\Sigma-D$ relation, instead of the classical vertical regression. \citet{iso90} show that regression in astrophysical applications is more complex than most realize and give alternatives to ordinary vertical regression. In our paper, following the
conclusions of \citet{iso90}, \citet{green05} and \citet{ur2010}, we test four different linear regression methods (including already mentioned orthogonal
regression) that treat variables $\Sigma$ and $D$ symmetrically.
Symmetry between $\Sigma-D$ and $D-\Sigma$ is also required due to the existence of large scatter in both $D$ and $\Sigma$ axes.
 Therefore, the values of parameters obtained by fitting $\Sigma-D$ and $D-\Sigma$ relations have to agree within estimated uncertainties.

This paper also considers possible dependence of radio luminosity on linear diameter ($L-D$ dependence). This
criterion was established by \citet{arbo04}, and is as follows:
if $L-D$ relation is obtained then $\Sigma-D$ relation follows and it
may be used for the estimation of SNR distances.

In Section 2, we present a new theoretical derivation of the $\Sigma-D$ relation based on equipartition
between cosmic rays and magnetic field energy\footnote{By ''equipartition'' we do not
assume that the energies are necessarily equal or nearly equal, but rather that
the energy ratio is constant during evolution.}.
The equations derived here describe three consecutive phases of the adiabatic (Sedov) phase of evolution for non-thermal
radiation from SNRs. In Section 3, of the 274 SNRs in Green's present catalog (2009 March version), we construct a calibration sample containing
60 Galactic shell remnants with distance estimates, and use it to derive $\Sigma-D$ relation. Additionally, Monte Carlo simulations
are performed in Section 4 to estimate the influence of data scatter on the $\Sigma-D$ and $L-D$ slopes.
In Section 4, we also derive the empirical Galactic $\Sigma-D$ relation for the calibration sample containing 60 SNRs.
In Section 5, we apply our Galactic $\Sigma-D$ relation to obtain the distances of two newly discovered, large and faint SNRs, G25.1$-$2.3 and G178.2$-$4.2.
In Section 6, we present $\Sigma-D$ relations for a subsample of 28 Galactic SNRs evolving in dense environments and for a subsample only 5 Galactic SNRs evolving in low density. In Section 7, we briefly review selection effects which influence Galactic samples of SNRs severely. The conclusions of this paper are presented in the last section.

\section{A theoretical interpretation of the $\Sigma-D$ relation}

A slightly different theoretical interpretation of the $\Sigma-D$ relation is presented for non-thermal radiation from SNRs. This model
is based on equipartition introduced by \citet{reyche81}.
The equipartition is a useful tool for estimating magnetic field strength and
energy contained in the magnetic field and CR particles using only the radio synchrotron emission of a source.
Details of equipartition and revised equipartition calculations for radio sources
in general are available in \citet{pach70}, \citet{gif04} and \citet{bik05}, respectively.
\citet{arbo12} introduced new modifications in equipartition calculation. They integrated
over momentum to obtain energy densities of particles so there is no need to introduce a break in the
energy spectrum and also take into account different ion species.

The total energy density of cosmic rays (CRs) can be written as the sum of two components \citep{arbo12}: energy of electrons ($\epsilon_{\rm{e}}$) and
ions ($\epsilon_{\rm{ion}}$)

\begin{eqnarray}
\epsilon_{\rm{CR}} = \epsilon_{\rm{e}} + \epsilon_{\rm{ion}} \approx \kappa K_{\rm{e}} (m_{\rm{e}}c^2)^{2-\gamma} \Bigg[ \frac{1}{\sum _i Z_i \nu _i}
\Big(\frac{m_{\rm{p}}}{m_{\rm{e}}}\Big)^{2-\gamma} \cdot \Big(\frac{2 m_{\rm{p}} c^2 E_{\rm{inj}} }{E_{\rm{inj}}^2 + 2 m_{\rm{e}}c^2 E_{\rm{inj}}} \Big)^{(\gamma-1)/2}
\cdot \nonumber \\
\frac{\Gamma(\frac{3-\gamma}{2}) \Gamma(\frac{\gamma-2}{2})}{2 \sqrt{\pi} (\gamma-1)} \sum _i A_i ^{(3-\gamma)/2}\nu _i  \Bigg],
\label{total-energy}
\end{eqnarray}

\noindent where $\gamma$ is the energy spectral index ($2 < \gamma < 3$), $E_{\rm{inj}}$ is injection energy of particles, $m_{\rm{e}}$ and $m_{\rm{p}}$ are electron and proton masses, $\nu _i = n_i/n$ are ion abundances, $A_i$ and $Z_i$ are mass and charge numbers of elements, and $K_{\rm{e}}$ is the constant in the power-law energy distribution for electrons.
$\kappa$ is a slowly varying function (if $E_{\rm{inj}}$ is not very high, see
Fig. 2 in Arbutina et al. 2012), which incorporates the ratio between
electron and ion energy (the latter of which should be dominant).
Here, the term "total energy density" implies that we take into account all CR species (e.g. electrons, protons, $\alpha$-particles, heavier ions)
inside the SNR, which have been injected into the acceleration process.
We neglect energy losses.

We assume that an isotropic power-law distribution of ultra-relativistic electrons is created at the shock wave \citep{bell78}
\begin{equation}
N(E)dE = K_{\rm{e}}E^{-\gamma}dE .
\label{power-law}
\end{equation}

\noindent For a given element of gas, $K_{\rm{e}}$ evolves behind the shock wave as a result of conservation of energy. The flux density
of synchrotron radiation of ultra-relativistic electrons, obtained from \citet{pach70} after substituting the emission coefficient  $\varepsilon_{\nu}$ with flux density $S_{\nu}$, is

\begin{equation}
S_{\nu}  \propto K_{\rm{e}} B^{1+\alpha} V \nu^{-\alpha} \;\;\; \rm{\frac{W}{{m}^{2} \, Hz}},
\label{flux-dens}
\end{equation}

\noindent where $B$ is the magnetic field strength, $V$ is the volume, $\nu$ is the frequency and $\alpha$ is the synchrotron spectral index
defined as $\alpha = (\gamma-1)/2$. Here, we use the flux density defined as

\begin{equation}
S_{\nu} = \frac{L_{\nu}}{4\pi d^{2}} = \frac{\varepsilon_{\nu} V}{d^{2}} = \frac{4\pi}{3} \varepsilon_{\nu} f \theta^3 d,
\label{flux-dens-def}
\end{equation}

\noindent where $L_{\nu}$ is the radio luminosity, $f$ is the volume filling factor of radio emission,
$d$ is the distance and $\theta = R/d$ is the angular radius. Then, we obtain the $\Sigma-D$ relation

\begin{equation}
\Sigma_{\nu} = AD^{-\beta} = \frac{L_{\nu}}{\pi^{2} D^{2}} \propto S_{\nu} D^{-2} \;\;\;  \rm{\frac{W}{{m}^{2} \, Hz \, sr}}.
\label{flux-dens}
\end{equation}

\noindent where $\Sigma_{\nu}$ represents surface brightness defined as $\Sigma_{\nu} = S_{\nu} /  \Omega$ where $\Omega$ (in steradians) is the solid
angle of the radio source.

\citet{reyche81} assumed redistribution of energy between cosmic rays, magnetic field and thermal gas behind
shock wave (equipartition between $\epsilon_{\rm{CR}}$ and $\epsilon_{B}$) described with:

\begin{equation}
\epsilon_{\rm{CR}} \approx  \epsilon_{B} = \frac{1}{2 \mu_{\rm{0}}} B^{2} \propto \rho_{\rm{0}} \upsilon_{\rm{s}}^{2},
\label{equipartition}
\end{equation}

\noindent where $\upsilon_{\rm{s}}$ is the pre-shock velocity and $\rho_{\rm{0}}$ is the preshock ambient density. We introduce the parameter $s$
 to allow for an ambient density profile of the form $\rho \propto R^{-s}$,
$R \propto t^{m}$ and $\gamma = 2 \alpha + 1$, where $R$ represents radius ($D=2R$) and $m$ is known as a deceleration parameter \citep{bp04}.
This relation will also contain a break, similar to the relation of \citet{dju86}.

If we assume low shock velocity (older SNRs) so that $E_{\rm{inj}} = m_{\rm{p}} \upsilon_{\rm{s}}^{2} \ll 2 m_{\rm{e}} c^{2} \;$ i.e. $\; \upsilon_{\rm{s}} \ll 7000 \; \rm{km/s}$, from equation (\ref{total-energy}), we have

\begin{equation}
K_{\rm{e}} \propto \rho_{\rm{0}} \upsilon_{\rm{s}}^{2},
\label{Kstari}
\end{equation}

\noindent and, after expressing $B \propto (\rho_{0} \upsilon_{\rm{s}}^{2})^{1/2}$ from equation (\ref{equipartition}) and
 putting $V \propto D^{3}$, with $\upsilon_{\rm{s}}=\frac{dR}{dt} \propto mt^{m-1} \propto m D^{\frac{m-1}{m}}$ we find the relation

\begin{equation}
S_{\nu} \propto K_{\rm{e}}^{\frac{\gamma + 5}{4}} V \propto (\rho_{\rm{0}} \upsilon_{\rm{s}}^{2})^{\frac{\gamma + 5}{4}} D^3 \propto  D^{\Big( \frac{2(m-1)}{m}-s \Big) \frac{\alpha + 3}{2} + 3  } ,
\label{Sni}
\end{equation}

\noindent Taking $s=0$ and $m = 2/5$ \citep{sed59} leads to

\begin{equation}
\Sigma_{\nu} \propto D^{-\frac{3\alpha + 7 }{2}} ,
\label{SDstari}
\end{equation}

\noindent and substituting an average SNR spectral index $\alpha = 0.5$ gives $\beta = 17/4 = 4.25$ \citep[see][]{reyche81, bif04}.

Another possibility is $E_{\rm{inj}} \gg 2 m_{\rm{e}} c^{2} \;$ i.e.$\; \upsilon_{\rm{s}} \gg 7000 \; \rm{km/s}$, which is satisfied for younger remnants.
In this case, following equation (\ref{total-energy}), constant $K_{\rm{e}}$ in the power-law energy distribution for electrons has the form

\begin{equation}
K_{\rm{e}} \propto \rho_{\rm{0}} \upsilon_{\rm{s}}^{2} E_{\rm{inj}}^{\frac{\gamma - 1}{2}} \propto \rho_{\rm{0}} \upsilon_{\rm{s}}^{2(\alpha + 1)},
\label{Kmladi}
\end{equation}

\noindent and therefore the flux density is

\begin{equation}
S_{\nu} \propto \rho_{\rm{0}}^{\frac{\alpha + 3}{2}}  \upsilon_{\rm{s}}^{3(\alpha + 1)} D^3  \propto D^{ \frac{3(\alpha + 1)(m - 1)}{m} - \frac{s(\alpha + 3)}{2} + 3 }.
\label{Sni-2}
\end{equation}

Taking $s=0$ and $m = 2/5$ yields

\begin{equation}
\Sigma_{\nu} \propto D^{-\frac{9\alpha + 7 }{2}} ,
\label{SDmladi}
\end{equation}

\noindent and substituting $\alpha = 0.5$ gives $\beta = 23/4 = 5.75$.

\citet{bif04} applied time-dependent nonlinear kinetic theory for cosmic ray (CR) acceleration in SNRs for
studying the properties of the synchrotron emission. In particular, they applied detailed numerical calculations
for deriving the surface brightness-diameter ($\Sigma-D$) relation for the range of the relevant physical parameters, namely
ambient density ($n_{\rm{H}}$) and supernova explosion energy ($E_{\rm{sn}}$). \citet{bif04} derived ($\Sigma-D$) relations for different
phases of SNR evolution.

During the initial part of the free expansion phase, the expected dependence is

\begin{equation}
\Sigma_{\nu} \propto n_{\rm{H}}^{7/4} D ,
\label{free-exp1}
\end{equation}

\noindent which explains well the numerical $\Sigma_{\nu}(D)$ behavior for small $D$. In this region $\Sigma_{\nu}(D)$ does not depend on the
SN parameters $E_{\rm{sn}}$ and $M_{\rm{ej}}$ (mass of ejected matter), but does depend on the ISM density.

During the later part of the free expansion phase the expected dependence is

\begin{equation}
\Sigma_{\nu} \propto (E_{\rm{sn}}/M_{\rm{ej}})^{7/16} n_{\rm{H}}^{21/16} D^{-5/16}.
\label{free-exp2}
\end{equation}

In the subsequent Sedov\footnote{Also known as Sedov-Taylor phase} phase, we have

\begin{equation}
\Sigma_{\nu} \propto E_{\rm{sn}}^{7/14} D^{-17/4} ,
\label{sedov1}
\end{equation}

\noindent independent of $n_{\rm{H}}$ and $M_{\rm{ej}}$.

In the late Sedov phase, dependence goes toward

\begin{equation}
\Sigma_{\nu} \propto E_{\rm{sn}}  n_{\rm{H}}^{3/4} D^{-2}.
\label{sedov2}
\end{equation}

We note that the phase described by equation (\ref{SDmladi}), based on equipartition, represents an additional intermediate phase between
late free expansion and early Sedov phase introduced by \citep{bif04}. Therefore,  equations (\ref{SDmladi}), (\ref{sedov1}) and (\ref{sedov2}) respectively
describes three consecutive phases of the adiabatic (Sedov) phase of evolution of SNRs. SNRs spend most of the time of their
evolution in the Sedov phase, up to almost a million years in a hot ISM (McKee and Ostriker, 1977). Usually, pressure effects
are responsible for terminating the Sedov phase earlier.

Based on previous results and assuming that the majority of SNRs are in the Sedov phase, we expect to obtain $\Sigma-D$ slope $\beta$ in the range of 2 to 5.75 depending on the evolutionary stage of the remnants in
a sample of shell-type SNRs. Also, we should expect additional scatter (which does not affect the slope significantly) in the $\Sigma-D$ diagram due
to different properties of the SN explosion and ISM, such as the energy of explosion, the mass of the ejected matter and the density of the ISM.

\section{Calibration sample and new $\Sigma-D$ relation}

To obtain a reliable empirical $\Sigma-D$ relation, we used a sample of shell-type SNRs
with independently determined distances and well-determined angular diameters and
flux densities. Various methods provide distance estimates to the remnants such as
proper motions, shock and radial velocities, HI absorption and polarization, association or interaction with HI, HII and CO molecular clouds,
X-ray observations, optical extinction and low frequency radio absorption.
Of the 274 SNRs in Greens's present catalog (2009 March version), 84 have
independently determined distances (59 of them classified as shell type, 14 composite, 6 filled center and
5 unknown type). The flux densities given in the catalog and the calculated surface brightnesses in this paper are referenced to 1 GHz.

The surface brightnesses and angular diameters are taken from Green's catalog. This catalog was also our primary source of information
about the distances of remnants. Many of the SNRs from this catalog have more than one distance available.
Following CB98 approach, we have either chosen the most recent measurement
or used an average of the available estimates (if the given distance range is narrow enough). We have also searched the literature for
 recent papers not included in Green's catalog, which provide accurate distances to Galactic shell remnants. Our Galactic sample of 60 shell remnants
with direct distance estimates, which is used to derive new $\Sigma-D$ relation, is listed in Table 1.

Part of our Galactic sample contains 33 SNRs, with updated parameters, which were also used as calibrators by CB98. The following 4 SNRs from the CB98 sample were omitted from our sample. According to Green's catalogue, the shell structure of G49.2-0.7 (W51) is questionable, as well as flux
density at 1 GHz, while G330.0+15.0 (Lupus Loop) does not have enough accurate measurements of angular diameter and flux density. Up to now, only a lower limit for distance of remnant G304.6+0.1 (Kes 17) has been measured (9.7 kpc from HI absorption). \citet{fos06} showed that OA 184, previously classified as SNR G166.2+2.5, is actually a Galactic HII region energized by O7.5V star BD+41$^{\circ}$1144.

Our Galactic sample also includes 5 composite SNRs because: (i) they have pure shell structure in radio regardless of centrally-brightened radio morphology; or (ii) it was possible to separate the shell flux density from the pulsar wind nebula (PWN) flux density. \citet{tam02} determined the integrated flux over various regions in the composite source G11.2-0.3. From their measurements and integrated flux density measurements of G11.2-0.3 at different frequencies, we estimate the SNR shell flux density at 1 GHz to be 20.5 Jy (subtracting PWN flux density which represents about 3 $\%$ of the flux density of the entire source). G93.3+6.9 (DA 530) is a high Galactic latitude SNR with a well-defined shell-like radio morphology which has a centrally
filled morphology only in X-rays \citep{jia07}. Observed in radio, the composite SNR G189.1+3.0 (IC 443) consists of two connected, roughly spherical shells of radio synchrotron emission, which are centered at different locations. Combining their measurements with existing data, \citet{cast11a} estimated that the flux density of the PWN represents only about 0.1$\%$ of total flux density and therefore we
calculated the SNR shell flux density at 1 GHz to be 164.7 Jy. \citet{cast11} investigated in detail the radio emission belonging to SNR G338.3-0.0 around the X-ray
pulsar candidate in a search for traces of a PWN, by reprocessing data corresponding to observations acquired with the ATCA \footnote{The Australia Telescope Compact Array (ATCA), an array of six 22-m antennas used for radio
astronomy}. No nebular radio emission has been found to correspond to the X-ray PWN, in spite of the good quality of
their radio images down to low surface brightness limits, and therefore the measured flux density in radio corresponds only to the SNR shell flux density.
Based on the radio images and the comparison with X-ray and IR observations, \citet{gia11} confirmed that there is no PWN within composite remnant
G344.7-0.1. Also, the same authors redetermined the distance of this SNR as (6.3 $\pm$ 0.1) kpc, on the basis of HI absorption and emission.

The shell remnant G1.9+0.3 is the youngest known Galactic SNR with known distance and the only Galactic SNR increasing in flux. Its estimated age is 156 $\pm$ 11 yr \citep{car11} and
distance about 8.5 kpc. Therefore, this SNR should not be included in our sample because it is probably still in the phase of free expansion.

We mentioned in Section 1 that in the past a single linear regression method was used for the purpose of obtaining the empirical $\Sigma-D$ relation: ordinary least-squares
regression of the dependent variable $Y$ against independent variable $X$, or OLS$(Y|X)$\footnote{Notation introduced by \citet{iso90}}. In OLS$(Y|X)$, the
regression line is defined to be that which minimizes the sum of the squares of the Y residuals.
Some applications, however, require using alternatives to OLS$(Y|X)$. The class of alternatives to OLS$(Y|X)$ used in our paper was suggested by
\citet{iso90} for problems where the intrinsic scatter of data dominates any errors arising from the measurement process.
This class of methods has also been proposed in order to avoid specifying "independent" and "dependent" variables.

The most important purpose of the $\Sigma-D$ relation is to estimate diameters (and hence distances)
for Galactic SNRs based on their observed surface brightnesses. \citet{green05} also stated an important issue related to the $\Sigma-D$ fits, even in the
case of disregarding problems with the selection effects. Namely, he pointed out that the measured value of $\Sigma$ is used to predict $D$, and that
a least squares fit minimizing the deviations in $\log{D}$ should be used to that effect. This, however, has not been done thus far,
and fits minimizing the deviations in $\log{\Sigma}$ have been used instead. He also mentioned that a fit to data
that treats $\Sigma$ and $D$ symmetrically is appropriate if a $\Sigma \propto D^{n}$ relation is used to describe the relationship between $\Sigma$ and $D$ and, by extension, the radio-evolution of SNRs.

Four methods that treat the variables symmetrically have been suggested by \citet{iso90}. One
is the line that bisects the OLS$(Y|X)$ and the inverse OLS($X|Y$) lines, called "OLS bisector"
or "double regression", which has been applied
in characterization of the Tully-Fisher and Faber-Jackson relations to estimate
galaxy distances \citep{rub80, lyn88, pie88}. The second method is the geometric mean of
OLS$(Y|X)$ and the OLS($X|Y$) slopes, proposed as the "impartial" regression by astronomer \citet{stro40} and used in
cosmic distance scale applications. It was derived also by statisticians, independently, and
called "reduced major-axis". The third regression is the line that minimizes the sum of the
squares of the perpendicular distances between the data points and the line, often called
"orthogonal regression" or "major-axis" regression. \citet{ur2010} applied orthogonal regression, for the first time,
in obtaining empirical  relation for SNRs in the starburst galaxy
M82. The fourth method that treats the variables symmetrically
represents the arithmetic mean of the OLS$(Y|X)$ and OLS$(X|Y)$ slopes \citep{aaro86}. It is easily recognized that these
four techniques, though each is invariant to switching variables (referred to as the dependent and independent),  lead to
completely different regression lines, both mathematically and in real applications.

\begin{figure}
\epsscale{1.0}
\plotone{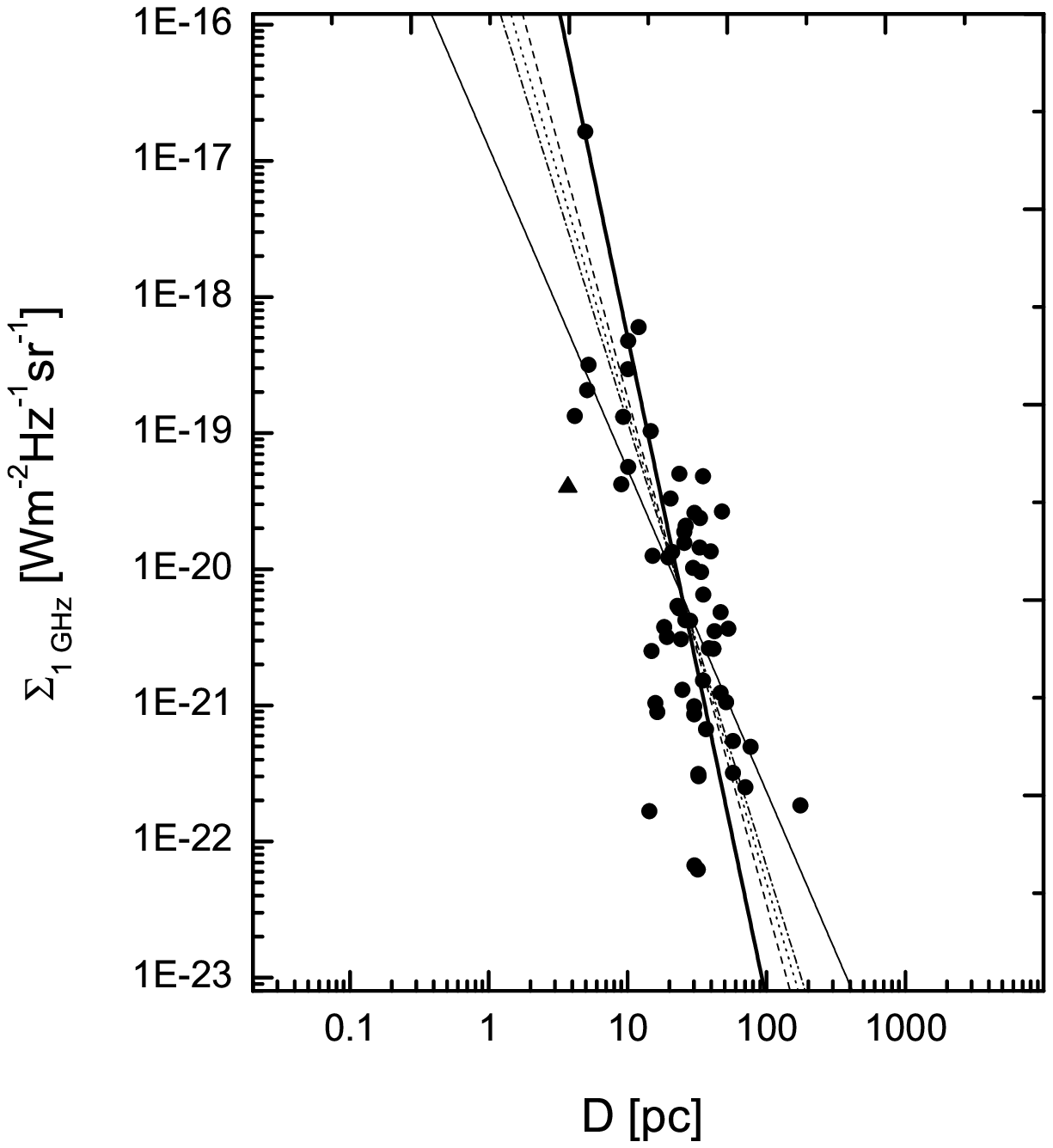}
\caption{The surface brightness vs. diameter $\Sigma-D$ relation at 1 GHz for shell
SNRs obtained by using the distance calibrators in Table 1. The different methods for minimizing the distance
of the data from a fitted line are presented.
Two solid lines represent OLS($Y|X$) (thin line) and orthogonal regression (thick line).
Dash, dot and dash-dot lines represent arithmetic, geometric mean of the OLS($Y|X$) and OLS($X|Y$) slopes and OLS bisector respectively. OLS($X|Y$)
line, with slope very similar to orthogonal regression, is omitted to avoid complicated graph.
G$1.9+0.3$, the youngest Galactic SNR which shows the flux density increasing with time \citep{green08}, is also shown (triangle) but not included in the calibration sample because it is still in the early (raising) free expansion phase of evolution \citep[see][]{bif04}.
\label{fig:SigmaD}}
\end{figure}

The empirical $\Sigma-D$ relation depends greatly on the regression method adopted. The dispersion of the six estimates
is considerably larger than the variance of any one estimate (see Table 2). For problems like these,
\citet{iso90} suggest to calculate all six\footnote{Actually, the sixth regression method,
 the arithmetic mean of the OLS$(Y|X)$ and OLS$(X|Y)$ slopes was not contained in their paper, but was added in proof.} regressions and to be appropriately cautious regarding the confidence of the inferred conclusion. Accordingly,
we give all mentioned regressions in Figure~\ref{fig:SigmaD} (except OLS$(X|Y)$, which is very close to orthogonal regression), applied to our calibration sample of 60 shell SNRs.

Data fitting is performed analytically and by using bootstrap. The fit parameter values and their errors,
presented in Table 2,
 were obtained after $10^6$ bootstrap data re-samplings for each fit. After applying bootstrap, we obtain the distribution of a
selected value for a certain regression procedure and then fit this dependence in OriginPro 8 with Gauss (normal) distribution, by
using 4-parameter fitting without parameter fixing. A normal distribution is used here as a good approximation to describe the behavior of bootstrap re-samples. Thus, we get mean (expectation) and associated standard deviation for each parameter.

\section{Monte Carlo Simulations}

\subsection{$\Sigma-D$ relation}

At first glance, inspection of Figure 1 and Table 2 leads to the conclusion that the resulting fit parameter values are
significantly influenced by the type of fitting procedure. This is mainly due to the large dispersion in the sample
of calibrators (low correlation coefficient\footnote{Pearson product-moment correlation coefficient is
calculated using the following equation:\\ $r=\frac{\sum_{i=0}^n (x_{i} - \bar{x})(y_{i} - \bar{y})}{\sqrt{\sum_{i=0}^n (x_{i} - \bar{x})^2} \sqrt{\sum_{i=0}^n (y_{i} - \bar{y})^2}} $.}
 $r=-0.68$). Much of this dispersion is due to properties of the SN explosion and ISM, which may
 substantially differ from one SNR to the other. It has generally been accepted that the density
 of the ISM is of significant importance in the evolution of SNRs; the other parameters are, in one way or another,
 connected to the ISM density \citep{ariur05}. SNRs of different types can be found along more or less
 parallel tracks in the $\Sigma-D$ plane. Also, the environment is probably
 quite inhomogeneous in the case of a single SNR, which would add more confusion in statistical studies.
 The errors of determined distances and, to a lesser extent, flux densities of the calibrator sample directly affect the scatter in the graph.

 Theoretical considerations assume that the $\Sigma-D$ relation corresponds to the evolutionary track of a typical SNR.
 Even for the Sedov phase, the fit should not necessarily be assumed to be linear (in log-log space) and Sedov sub-phases
 have $\beta$ slopes from 2 to 5.75 as shown in Section 2. The previously described effects adds an extra scatter in our sample
 of calibrators.

 Due to the lack of information, it is not possible to separate all SNRs according to their intrinsic
 properties, which are connected with density of ISM in which they expand and their evolutionary stage. We therefore, expect to
 obtain, in some sense averaged, empirical $\Sigma-D$ relation that may still be used as distance estimator
 in cases when other existing methods are inapplicable. Errors of distances by this method will not
 be negligible but are still acceptable for the above purpose (the average fractional error of the  CB98 relation defined
 as $f=\frac{d_{\rm{obs}}-d_{\rm{sd}}}{d_{obs}}$ was about 40\%).

 We performed a set of Monte Carlo simulations to estimate the influence of scatter of the data
 on the $\Sigma-D$ slope. We generated random SNR populations (50 to 5000 SNRs) according to artificially
 chosen $\Sigma-D$ relations with $\beta$ slopes 2, 3, 4 , 5 and 6, covering roughly the expected
 interval of slopes in Sedov phase. For each randomly selected value of $D$ from interval 5 to 150 pc (similar to that of real data)
 we calculate surface brightness $\Sigma$ using the initially proposed relation. Then we add random "polar" scatter in the created
 log-log data set. We add scatter by choosing random angle $\phi$ and random distance $r$ from original point $(x,y)$, after
 which we perform the following transformation $ (x,y) \longmapsto (x+r\cos{\phi},y+r\sin{\phi})$. For obtaining random angle $\phi$ and
 distance $r$, respectively, we use uniform distribution and $\chi^2$ distribution with 10 degrees of freedom (usually labeled as $\chi_{10}^2$)
 implemented in GSL\footnote{GNU Scientific Library (GSL), \url{http://www.gnu.org/software/gsl/}} numerical library for C programming language. We have chosen this distance
 distribution to be close to normal distribution, although additional tests showed that the type of distribution and number of degrees of freedom (for $\chi^2$) does not significantly affect our later conclusions. The only parameter which we enter is the distance that occurs most frequently in a generated
 distribution of distances (known as the mode in statistics)
 and therefore, it is a direct measure of dispersion from generated artificial samples (denoted by $r_{\rm{mode}}$). We have chosen such a mode value which ensures that the artificial distribution is similar to the real distributions for the 60 SNRs sample fitted with six different regression methods ($r_{\rm{mode}} \approx 0.5$).

\begin{figure}
\epsscale{1.0}
\plotone{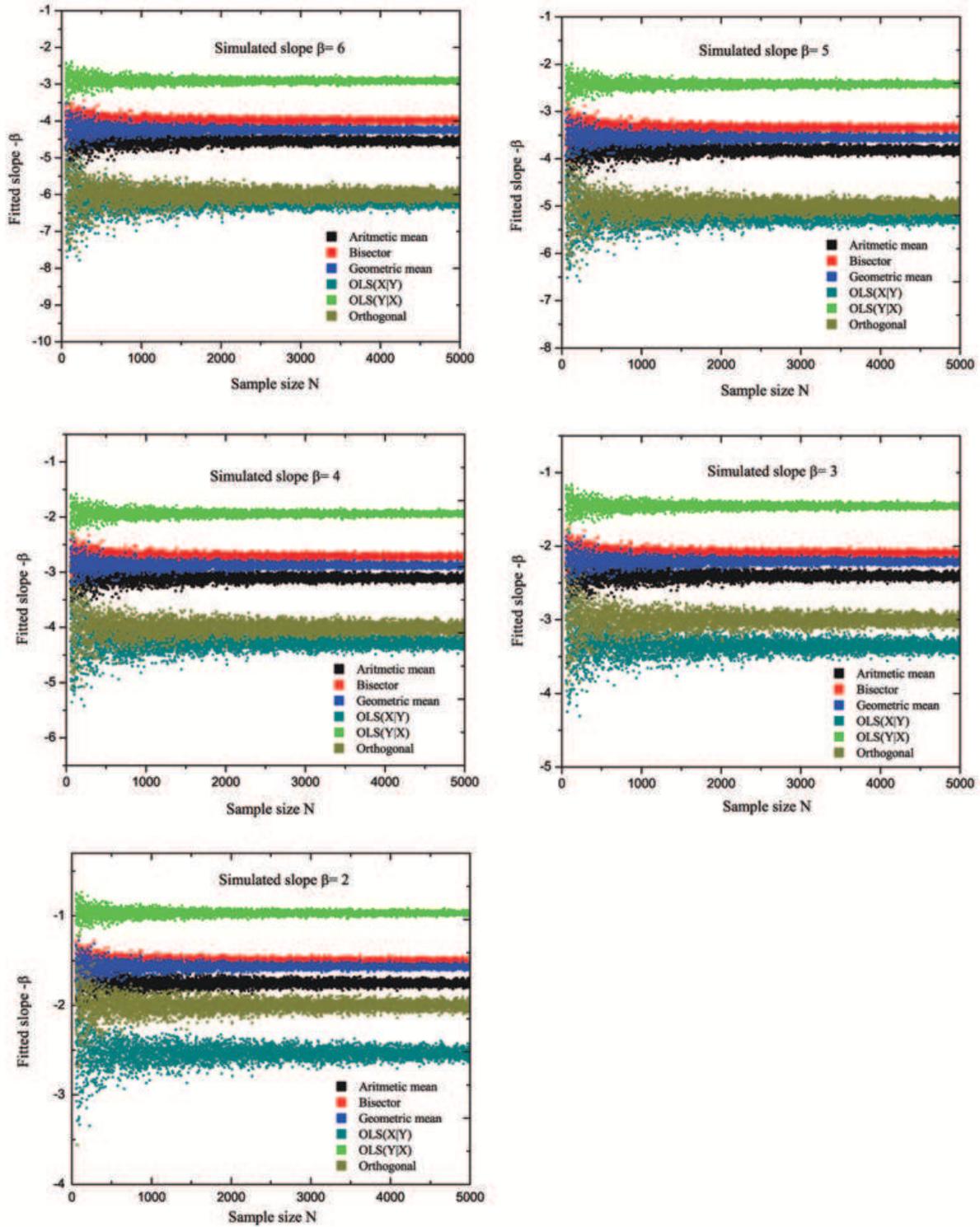}
\caption{The fitted $\Sigma-D$ slopes vs. sample sizes for six regression types applied on randomly generated SNR samples containing artificial scatter
and simulated slopes $\beta$=2, 3, 4, 5, 6. This analysis was done to determine which type of regression is the
least sensitive to scattering.  \label{fig:SDuporedni}}
\end{figure}

 We give values of fitted slopes for randomly generated SNR populations in Figure~\ref{fig:SDuporedni} as a function of sample size and
 regression method (orthogonal, arithmetic mean, geometric mean, bisector, OLS$(Y|X)$ and OLS$(X|Y)$ regression). Each figure also
 contains information about the value of simulated slope $\beta$ before performing random scatter on the log-log data set.

 After inspection of graphs on Figure~\ref{fig:SDuporedni} we come to the conclusion that orthogonal regression gives $\beta$ slopes which are
 closest to the initial slopes of the artificially generated samples. Also, this type of regression shows significant stability as it
 converges to narrow value interval for larger sample sizes.

 This means that orthogonal regression is least
 sensitive to the scatter in the $\Sigma-D$ plane. As mentioned at the beginning of this section, data dispersion is large in the calibration
 set that is used to construct our new empirical Galactic $\Sigma-D$ relation. Although \citet{iso90} concluded that the bisector performs
 significantly better than orthogonal regression, our class of problems (with large intrinsic scatter and steep
slopes) gives more reliable solutions when solved with orthogonal
regression. The existence of
this significant scattering is obvious from the plots. It occurs as a
result of coupling of several intrinsic SNR properties related to energy
liberated by supernova explosions, density of surrounding media and
evolutionary status of the remnants.
The results of our Monte Carlo simulations (see Figure 2) show that the orthogonal regression give fitted slopes that are closest in value to the simulated slopes.
Therefore, we strongly suggest orthogonal regression, instead of the previously used OLS$(Y|X)$ and other regression types, for obtaining
 any kind of empirical $\Sigma-D$ relation with slopes between 2 and 6. All previous Galactic empirical relations have slopes in this interval (see \citealt{ur2002}).

 After applying non-weighted orthogonal regression on the sample containing 60 calibrators from Table 1, we obtain the relation

\begin{equation}
\Sigma_{\rm{1 GHz}} = 3.25^{+27.94}_{-2.91} \cdot 10^{-14}  D^{-4.8 \pm 0.7} \hspace{0.3cm} \textnormal{W} \textnormal{m}^{-2} \textnormal{Hz}^{-1} \textnormal{sr}^{-1}.
\label{relacija1}
\end{equation}

\noindent The new relation is significantly steeper than those obtained in earlier studies.
The results of our Monte Carlo simulation
(see Figure 2) also gave such a steep slope for the examined relation.
Moreover, the orthogonal fit gives steeper slopes than any other
regression, except OLS$(X|Y)$.
Shell SNRs with distances derived from our $\Sigma-D$ relation are shown in Table 3. In total, 207 remnants are shown.

Next, we define fractional errors
\begin{equation}
f = \left|  \frac{d_{\rm{I}} - d_{\Sigma}}{d_{\rm{I}}}  \right|
\label{fractional}
\end{equation}
\noindent in order to get an estimate of the accuracy of the $\Sigma-D$ relation for
 individual SNR distances and also as an indicator of the applicability of our
 relation for distance determination. Here $d_{\rm{I}}$ is the independently determined distance to
 an SNR and $d_{\Sigma}$ is the distance derived from the $\Sigma-D$ relation.
 Here, the average fractional error is $ \bar{f}=0.47$ (comparable to that
 of CB98 $\bar{f}=0.41$, although we used a significantly larger sample). We also give average
 fractional errors for all six regressions in Table 2.

The Monte-Carlo simulations presented here show that orthogonal regression is more stable for larger samples (see Figure 2), so our calibration data-set (60 SNRs) does not represent a proper sample for obtaining a high accuracy $\Sigma-D$ relation. A more stable behavior for larger samples is also the case for the other five regressions.

\subsection{$L-D$ relation}

\begin{figure}
\epsscale{1.0}
\plotone{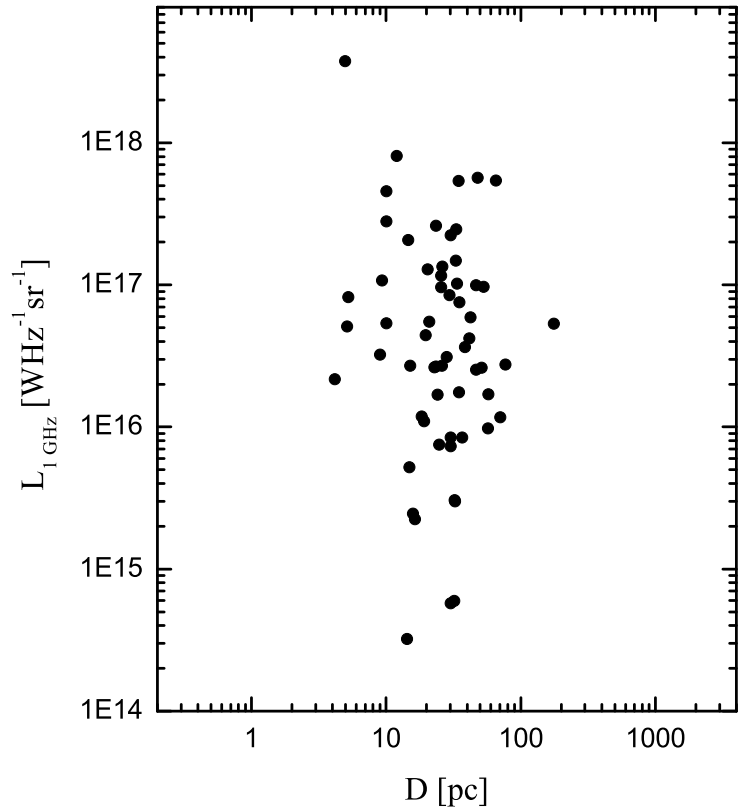}
\caption{The radio luminosity (at 1 GHz) vs. the diameter for the SNRs ( $\Sigma-D$ calibrators) given in Table 1,
to allow a possible dependence of the luminosity on linear diameter in form $L_{\nu}=CD^{\delta}$.
Weak correlation is evident, primarily due to severe data scatter.
\label{fig:LD}}
\end{figure}

For a proper $\Sigma-D$ analysis, the $L-D$ correlation should be checked. Using appropriate definitions of flux density and angular diameter, we have the
following dependence $\Sigma_{\nu} \propto L_{\nu} D^{-2} $.
However, the $\Sigma-D$ relation can be written as

\begin{equation}
\Sigma_{\nu}= A D^{-2 + \delta},
\label{LD2}
\end{equation}

\noindent to allow for a possible dependence of luminosity on the linear diameter in the form $L_{\nu}=CD^{\delta}$, where $C$ is constant.
Radio luminosity as a function of the diameter for the calibrators in Table 1 is given in Figure~\ref{fig:LD} and a
very low correlation is evident. As an illustration, orthogonal regression, applied here, gives $L-D$ slope $\delta \approx -15.3$.
However, does not have a physical meaning because $\delta$ should be around -3, according to equation~(\ref{LD2}).
Also, CB98 did not find any significant correlation between luminosity
and linear diameter for shell type alone or for
shell $+$ composite-type distance calibrators.
Previous studies of the $\Sigma-D$ relation indicate that $\delta \approx 0$, which leads to a radio luminosity that is independent
of diameter and to the so-called trivial relation, $\Sigma \propto D^{-2}$. If the $L-D$ correlation does not exist, the
trivial $\Sigma \propto D^{-2}$ form should not be used \citep{arbo04}.

Similar to the previous subsection, we also perform a set of Monte Carlo simulations to estimate the influence of the severe scatter of
data on the $L-D$ slope. We use a randomly generated $\Sigma-D$ sample with added random polar scatter to construct $L-D$ scatter by
applying the luminosity-surface brightness relation $L_{\nu} = \pi^2 D^2 \Sigma_{\nu}$.

\begin{figure}
\epsscale{1.0}
\plotone{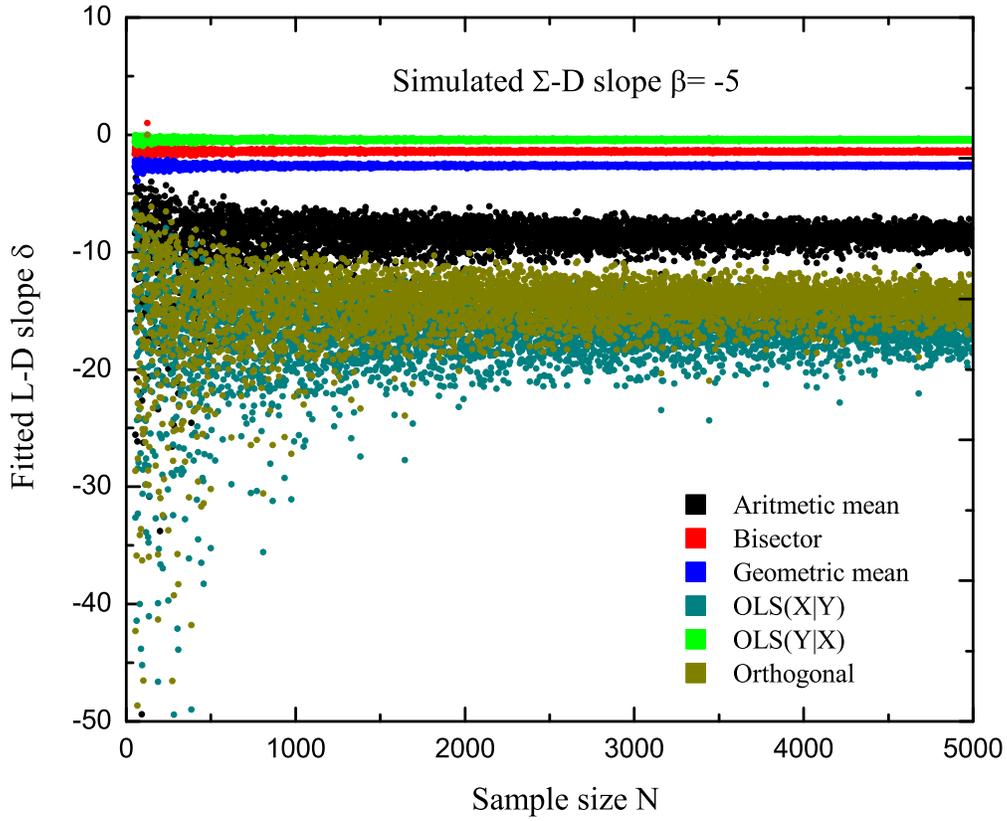}
\caption{The fitted slopes of the power-law fit $L_{\nu} = CD^{\delta}$ vs. sample sizes for six regression types
 applied on randomly generated samples ($r_{\rm{mode}}=0.5$). It is evident that
 the orthogonal fit becomes very
unstable for $L-D$ relation and strongly influenced by data scatter, especially for a small number of points,
as is the case with our sample. \label{fig:LD-uporedni}}
\end{figure}

Six regression methods were applied on randomly generated $L-D$ samples. Figure~\ref{fig:LD-uporedni} shows fitted $L-D$ slopes $\delta$ as
a function of sample size, with a scatter $r_{\rm{mode}}=0.5$ and the simulated $\Sigma-D$ slope $\beta=5$. At a first glance,
Figure~\ref{fig:LD-uporedni} leads to the conclusion that OLS$(Y|X)$, bisector and geometric mean regression (average $\delta$ values respectively
$-0.42, -1.41, -2.64$) give much flatter slopes than arithmetic mean, orthogonal and OLS$(X|Y)$ regressions ($\delta$ respectively $-8.84, -14.90, -17.25$).
We note that orthogonal regression, which is of significant importance for us, gives a very similar slope ($-14.9$) for randomly scattered data as
that for real data ($-15.3$), which is very good agreement but with no physical importance. The orthogonal fit performs very unstably, especially for
lower sample sizes ($N < 500$) and it is strongly influenced by data scatter, as are the other five fitting methods. Although OLS$(Y|X)$, bisector
and geometric mean regression seem more stable than other three methods, converting their slopes into angle interval leads to the opposite
conclusion.

\begin{figure}
\epsscale{1.0}
\plotone{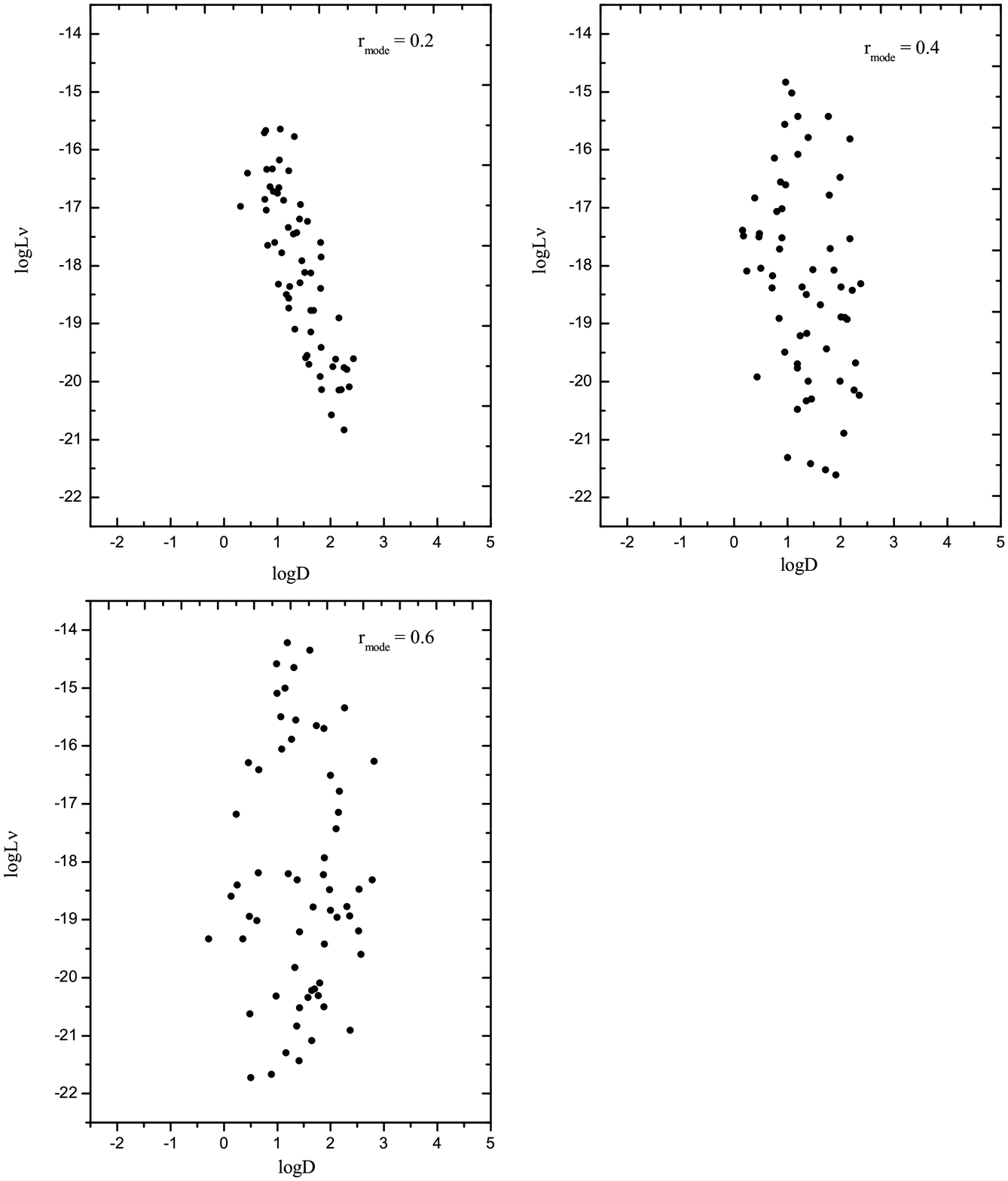}
\caption{Appearance of $L-D$ relation when increasing data dispersion (quantified with parameter $r_{\rm{mode}}$).
It can be inferred from the plots that larger scattering leads to the decrease and even loss of the correlation.
\label{fig:LDnestab}}
\end{figure}

The main reason for this instability can be inferred from brief inspection of Figure~\ref{fig:LDnestab}. When we increase the scatter of the
simulated data sets, the distribution of points tends to be almost vertical and even severe scatter can change the fitted slope from negative to positive.
This effect produces a break in the dependence of the orthogonal fit slope from data dispersion ($r_{\rm{mode}}$). This may be due to a lower span
of luminosities for the $L-D$ relation in comparison to the span of surface brightness for $\Sigma-D$ relation (spans of diameters are the same for both relations).

\begin{figure}
\epsscale{1.0}
\plotone{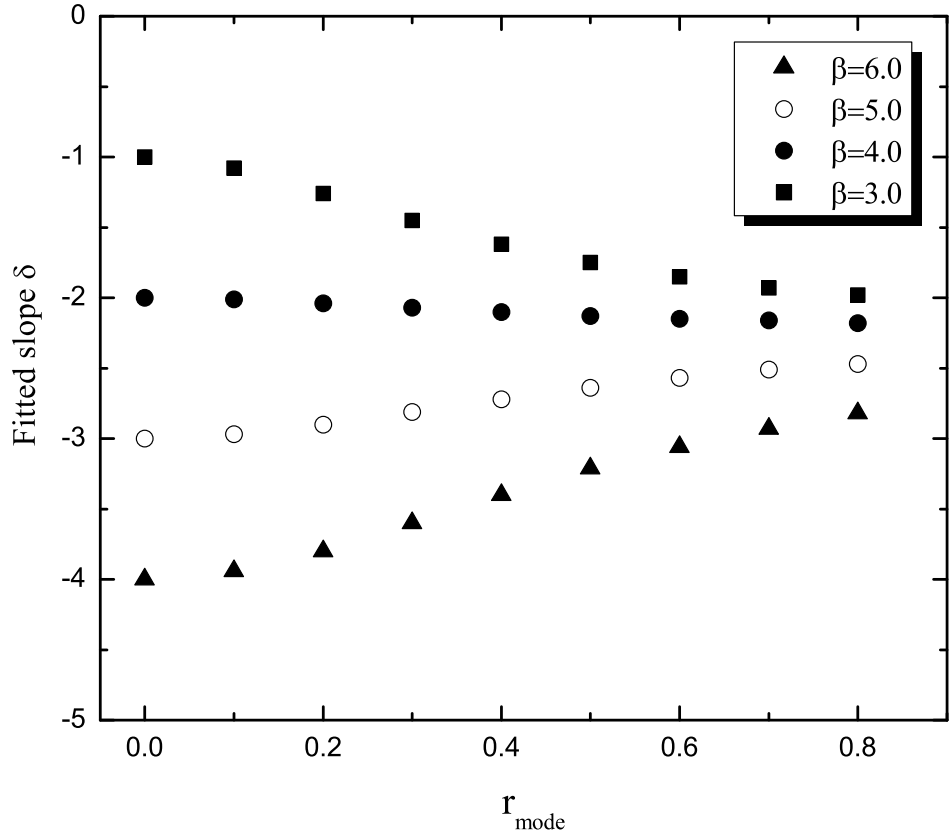}
\caption{Performance of geometric mean regression for fitting power-law $L-D$ relation $L_{\nu} = CD^{\delta}$.
This type of regression can be a valid tool for obtaining $L-D$ relation only for a limited range of data dispersion
and it is closest to theoretical value $\delta = -\beta + 2$.
\label{fig:Geometric}}
\end{figure}

It can be inferred from further analysis that, taking into account all six fitting procedures, the geometric mean regression gives a slope that is close
 to that expected $\delta = -\beta + 2$ (for simulated $\Sigma-D$ slope $\beta=6, 5, 4, 3$). Figure~\ref{fig:Geometric} also shows that this type
of regression can be a valid tool for obtaining $L-D$ relation for a limited range of data dispersion. Thus, we should be careful if geometric mean
regression is applied on $L-D$ analysis since it is not insensitive to severe scatter like orthogonal regression when applied on $\Sigma-D$
relation.

Therefore, we suggest to obtain the $L-D$ relation directly by using some type of regression only if the data set of calibrators is not subject to
 severe scatter. In other cases, a more accurate approach would first require the derivation of $\Sigma-D$ relation
 from which then, by using the equation (\ref{LD2}), we obtain the $L-D$ relation. Our conclusion is also supported by the fact that
 \citet{arbo04} did not find any significant correlation when analyzing $L-D$ dependence, except for the M82 starburst galaxy (with a fit quality of 64 per cent) for which a good $\Sigma-D$
relation does exist.

\section{Distances of two newly found Galactic SNRs: G25.1$-$2.3 and G178.2$-$4.2}

The Sino-German $\lambda$6 cm Galactic plane survey is a sensitive survey with the potential to detect new low-surface-brightness SNRs.
This survey searches for new shell-like objects in the $\lambda$6 cm survey maps and studies their radio emission, polarization and spectra
using the $\lambda$6 cm maps together with the $\lambda$11 cm and $\lambda$21 cm Effelsberg observations.

\citet{gao11} have discovered two new, large, faint SNRs, G25.1$-$2.3 and G178.2$-$4.2, both of which show shell structure.
G25.1$-$2.3 is revealed by its strong southern shell, which has a size of 80$^{\prime}$ $\times$ 30$^{\prime}$. It has a non-thermal radio spectrum with a spectral index
of $\alpha=-0.49 \pm 0.13$. G178.2$-$4.2 has a size of 72$^{\prime}$ $\times$ 62$^{\prime}$ with strongly polarized emission being detected along its northern shell.
The spectrum of G178.2$-$4.2 is also non-thermal, with an integrated spectral index
 of $\alpha=-0.48 \pm 0.13$. Its low surface brightness makes G178.2$-$4.2 the second faintest known Galactic SNR. This
demonstrates that more large and faint SNRs exist, but are very difficult to detect.

Using the total intensity and the radio spectral index, \citet{gao11}
calculated the surface brightness of both objects at 1 GHz, obtaining $\Sigma_{\rm{1 GHz}}$ = 5.0 $\cdot$ 10$^{-22}$ Wm$^{-2}$Hz$^{-1}$sr$^{-1}$ for
the southern shell of G25.1$-$2.3 and $\Sigma_{\rm{1 GHz}}$ = 7.2 $\cdot$ 10$^{-23}$ Wm$^{-2}$Hz$^{-1}$sr$^{-1}$
for G178.2$-$4.2.

The $\Sigma-D$ relation of CB98 was used to estimate the distances of the two newly found SNRs.
For G25.1$-$2.3, they found that the diameter for the shell is 72 pc and the distance is 3.1 kpc,
consistent with those derived from the HI data.
For G178.2$-$4.2, they found that the diameter is 197 pc and
its distance is 9.4 kpc. This places the object far outside the
Galaxy, which seems hardly possible.

This is the appropriate place to mention specific behavior of the $\Sigma-D$ relation, depending on the brightness
of the SNR for which the distance is calculated. The slope of the $\Sigma-D$ relation directly affects derived diameters and
hence distances. A steeper slope will give larger diameters (and hence distances) for SNRs with higher surface brightnesses, while
it will lead to smaller diameters (distances) for SNRs with lower surface brightnesses.

We applied our new Galactic $\Sigma-D$ relation to obtain the distances of these two objects.
For G25.1$-$2.3, we obtained that the diameter for the shell is 42 pc and hence the distance is 1.8 kpc,
which is still not far from that derived from the HI data.
For G178.2$-$4.2, we found that the corresponding diameter is only 63 pc, so that
its distance is 3.0 kpc, which places the object inside the
Milky Way.
Note, however, that the uncertainties of these estimates could be as large as 50\%.

\section{Dependence of $\Sigma-D$ relation on the density of the interstellar medium}

We focus on a more direct connection between the ISM density
and the $\Sigma-D$ relation.
\citet{dju86} assumed that the dependence of surface brightness on the density of the ISM has
the form $\Sigma \propto \rho_{0}^{\eta} \propto n_{\rm{H}}^{\eta}$ where
$\rho_{0}$ and $n_{\rm{H}}$ are the average ambient density and hydrogen number density,
respectively, and $\eta$ is a constant \citep[see also][]{bif04}.
This means, the larger the surrounding ISM density, the greater the synchrotron emission
from the SNR.

This suggests that, on average, SNRs in dense environments would tend to have higher surface brightness in comparison to those evolving in
lower-density medium. However, since $\Sigma = AD^{-17/4}$ relation depends primarily on the initial explosion energy (equation \ref{sedov1}),
variation in SN energy rather than density could produce scattering in some parts of the $\Sigma-D$ diagram.

Although it is difficult to decide which SNRs should be included or
excluded from analysis, we tried to extract subsamples of Galactic
SNRs (using Table 1) based on the density of the medium in which they evolve and then to obtain particular $\Sigma-D$
relation.

\citet{math83} classified SNRs into four categories based on their
optical characteristics: Balmer-dominated, oxygen-rich, plerion/composite and evolved SNRs.
Balmer-dominated line emission is produced when the expanding
SNR encounters low-density ($\sim 0.1 - 1 \rm{cm}^{-3}$), partially neutral ISM.
The hydrogen overrun by the shock is collisionally excited in a thin ($\lesssim 10^{15} \rm{cm}$) ionization zone, producing optical
spectra dominated by the Balmer lines of hydrogen \citep{gha07}. Oxygen-rich SNRs were mainly identified based on their optical
oxygen line emission properties and they primarily occur in HII and molecular cloud regions, i.e. in higher density regions.
Young oxygen-rich SNRs are likely to interact with especially complex circum-stellar medium, rather than ISM.
Balmer-dominated SNRs are connected to Type Ia SNe - deflagration of C/O white dwarf, while oxygen-rich SNRs originate in the
Type Ib events - explosions of a massive O or a Wolf-Rayet (W-R) star.

We created a subsample of 28 Galactic SNRs with known distances, using online database as part of paper\footnote{This paper builds
on the List of Galactic SNRs Interacting with Molecular Clouds maintained by Bing Jiang
\url{"http://astronomy.nju.edu.cn/~ygchen/others/bjiang/interSNR6.htm"}} \citet{gile12}, interacting with molecular
clouds, i.e. evolving in dense ISM. The properties of these SNRs are given in Table 4.
After applying orthogonal regression on the subsample from Table 4, the $\Sigma-D$ relation obtained is
\begin{equation}
\Sigma_{\rm{1 GHz}} = 3.89^{+12.81}_{-2.98} \cdot 10^{-15}  D^{-3.9 \pm 0.4} \hspace{0.3cm} \textnormal{W} \textnormal{m}^{-2} \textnormal{Hz}^{-1} \textnormal{sr}^{-1},
\label{relacija2}
\end{equation}
\noindent which is slightly steeper than relation obtained by \citet{arbo04} $\beta=3.5 \pm 0.5$, whose sample contained
only 14 Galactic SNRs associated with molecular clouds. The average fractional error is $ \bar{f}=0.35$. Slope $\beta=3.9$ for Galactic SNRs in dense ISM is in very good agreement with slope obtained
 for the M82 data sample by \citet{ur2010}, who also used orthogonal fitting.
 Our obtained slope approximately coincides with the conclusion of \citet{bif04} that
SNRs in dense ISM should populate the part of the line $\Sigma=AD^{-17/4}$.

\begin{figure}
\epsscale{1.0}
\plotone{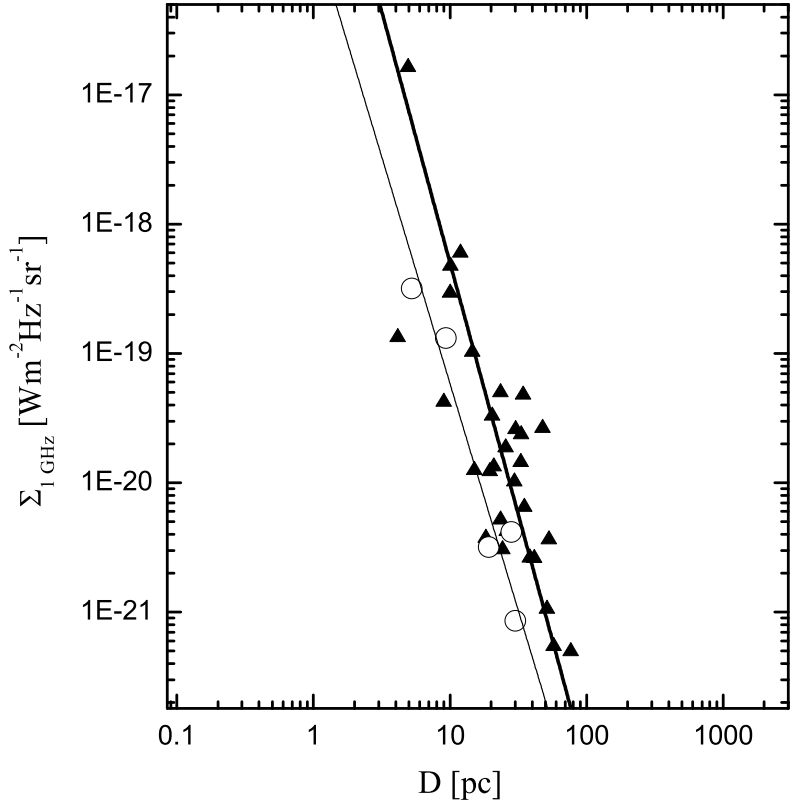}
\caption{ The $\Sigma-D$ plots at 1 GHz for Balmer-dominated (empty circles) and oxygen rich SNRs
(triangles) -- thin line for the former and thick line for the later sample. The fit lines are
obtained by orthogonal offsets.
As expected, remnants evolving in a dense environment lie above those evolving in low density
medium due to higher synchrotron emission.
\label{fig:Gusti-retki}}
\end{figure}

The sample of Galactic SNRs evolving in low-density ISM is significantly smaller. Up to now, only 5 Galactic
SNRs showing Balmer-dominated characteristics have been reported \citep[see review papers][]{vink12,heng10} and are therefore thought to expand in low-density medium. The Balmer-dominated sub-sample contains: Kepler (G4.5$+$6.8), Cygnus Loop (G74.0$-$8.5), Tycho (G120.1$+$1.4),  RCW 86 (G315.4$-$2.3) and
SN 1006 (G327.6$+$14.6), whose basic properties can be found in Table 1. The $\Sigma-D$ relation obtained is
\begin{equation}
\Sigma_{\rm{1 GHz}} = 1.89^{+4.08}_{-1.29} \cdot 10^{-16}  D^{-3.5 \pm 0.5} \hspace{0.3cm} \textnormal{W} \textnormal{m}^{-2} \textnormal{Hz}^{-1} \textnormal{sr}^{-1}.
\label{relacija3}
\end{equation}
\noindent with an average fractional error $ \bar{f}=0.18$. Figure~\ref{fig:Gusti-retki} shows the relation for Balmer-dominated and oxygen-rich Galactic SNRs for comparison purposes. The two relations
have similar slopes, but the former is below the latter, as we would expect for SNRs in low-density environments. To further investigate the influence of the ISM density on the $\Sigma-D$ relation, we should use more complete samples, especially sample of Balmer-dominated remnants.

\section{Selection effects}

Identification of Galactic SNRs is always accompanied by
selection effects arising from difficulty in identifying (i) faint SNRs and (ii) small angular size SNRs \citep[e.g. CB98,][]{green91, green05, ur05, ur2010}.
A high enough surface brightness for SNRs is required in order to distinguish them from the background Galactic emission. Any Galactic radio survey
 is severely biased by observational constraints. The surface brightness limit seriously affects the completeness of all catalogues
 of Galactic SNRs. The derived flux densities of many SNRs are also poorly determined.
 Small angular size SNRs are also likely to be missing from current catalogues and calibration samples. Their small size may cause
wrong classification of SNR type and also lead to incorrect calculation of their surface brightness.

Additionally, Malmquist bias severely acts in the Galactic samples making them incomplete. This is a type of volume selection effect which
naturally favors bright objects in any flux-limited survey because they are sampled from a larger spatial volume. The result is a bias
against low surface-brightness remnants such as highly evolved old SNRs. Only the extragalactic samples are not influenced by the Malmquist bias
since all SNRs in the sample are at the essentially same distance.

\citet{ur2010} performed a set of Monte Carlo simulations to estimate the influence of the survey sensitivity selection effect on the
$\Sigma-D$ slope for the M82 sample (31 SNRs). An appropriate sensitivity cutoff is applied to the simulated
data points, taking into account only points above the given sensitivity line.
They concluded that the sensitivity selection effect does not have a major impact on the $\Sigma-D$ slope.

Technological advances in radio telescopes and X-ray instruments will greatly increase the number of known supernova remnants (SNRs),
lead to a better determination of their properties and, thus, reduce the influence of the mentioned selection effects.

\section{Summary and conclusion}

This paper presents a re-analysis of the theoretical and especially empirical Galactic $\Sigma-D$ relation and also the dependence of
this relation on the density of the ISM. Motivation was found in the observed property that the empirical $\Sigma-D$ relation
strongly depends on the used regression due to severe scatter in data samples.
In contrast to the standard least-squares (vertical) regression, we examine the behavior of six different types of regression. We put
emphasis on those which satisfy the requirement that the values of
parameters obtained from the fitting of $\Sigma-D$ and $D-\Sigma$ relations
should be invariant within estimated uncertainties i.e. treat $\Sigma$ and $D$ symmetrically (namely: orthogonal regression, bisector,
arithmetic and geometric mean regressions).

The catalog of known Galactic SNRs has grown in size significantly since the work of CB98, from 215 to 274 SNRs. The
number of SNRs with known distances has also increased. We included the latest distance updates, if available, to derive a new Galactic
$\Sigma-D$ relation, using a sample of 60 shell SNRs. We concluded from our tests that the size of the sample significantly influences the
stability of any type of regression, and that more observations are necessary to increase the statistical significance of the sample.

We present an additional modification to the theoretical $\Sigma-D$ relation for SNRs in the adiabatic expansion (Sedov) phase.
This modification is based on the equipartition introduced by \citet{reyche81}. Our modification induces a new sub-phase lying between
late free expansion and early Sedov phase having slope $\beta = 5.75$. Also, our theoretical slope ($\beta = 4.25$) for relatively older SNRs in Sedov phase is in agreement with that of \citet{bif04}.

We have performed an extensive series of Monte Carlo simulations to evaluate numerically how
well each of the six regression methods approximate the artificially proposed $\Sigma-D$ dependence
with added scatter. The orthogonal regression is the most accurate slope predictor in data sets with severe scatter as it
gives the slope closest to the original value. For other regression procedures, the fitted slopes of the $\Sigma-D$ relation are
seriously affected by data scatter. Standard (vertical) fitting or OLS$(Y|X)$, which was used in previous
papers considering empirical $\Sigma-D$ relation, leads to a significant change in slope when applied on highly dispersed
data sets. When the scatter increases, the standard fitting leads to flatter slopes close to trivial ($\beta \approx 2$).
Also, similar conclusions apply to OLS$(X|Y)$, bisector, arithmetic and geometric mean regression.
A possible reason may be the inherent property of $\Sigma-D$ data sets which are defined by a narrow range of diameters (about 1 order
of magnitude) in comparison to the wide range of surface brightness (approximately 5 orders of magnitude).
We recommend using orthogonal regression for obtaining any type of empirical $\Sigma-D$ relation.

Our obtained slope ($\beta = 4.8$) is significantly steeper than previous values, it is far from the trivial one ($\beta \approx 2.0$) and
agrees with theoretical predictions for the Sedov phase of SNR evolution. Since our calibration sample contains 60 SNRs, which probably have different explosion
energies, evolve in different ambient density and may be in different phases of SNR evolution, the relation $\Sigma \propto D^{-4.8}$ could represent an averaged evolutionary track for Galactic SNRs. As such, it could be used potentially for estimating the distances of SNRs in our Galaxy.
A relatively large data scatter can be partly explained by the above mentioned influences.

Additional analysis is carried to examine possible dependence of radio luminosity on linear diameter ($L-D$ relation) for SNRs.
Also, this analysis may provide an answer as to whether determination of SNR distances
on the basis of a $\Sigma-D$ relation is possible. We did not find any significant correlation for our sample consisting of 60 SNRs,
while orthogonal regression gives a very steep slope ($\beta \approx -15$) which has no physical meaning. Monte Carlo simulations have revealed that the
$L-D$ relation is very sensitive to severe scatter in the data, which is surely present in our sample.
Therefore, it is possible to obtain $L-D$ relation only if the set of calibrators is not subject to severe scatter. Otherwise, we obtain
very steep slopes without physical meaning. We recommend an indirect approach to obtaining an $L-D$ relation, which requires obtaining the $\Sigma-D$
relation first (by orthogonal regression).

We made an attempt to find a homogeneous subsamples of Galactic SNRs in low-density and dense
environments and to obtain the $\Sigma-D$ relations for two particular classes of
SNRs. Applying orthogonal regression to the sample (which contains 28 SNRs) made of SNRs which evolve in dense environment of molecular clouds
(including oxygen-rich SNRs),
leads to the slope $\beta=3.9$. Our sample contains twice as many SNRs as that of \citet{arbo04} and the obtained slope is
slightly steeper. Our slope for SNRs in dense ISM is in good agreement with the result obtained
by \citet{ur2010} for the sample of 31 SNRs in the M82 galaxy ($\beta = 3.9$). \citet{ur2010} proposed their relation to be used for estimating distances
to SNRs that evolve in a denser interstellar environment, with number density up to 1000 particles per cm$^3$. This agreement may be very
useful because this is the best sample for the $\Sigma-D$ analysis, consisting of a relatively
high number of very small and very bright SNRs from starburst galaxy M82.

In contrast to the sample evolving in dense environment, we still have a very small sample of SNRs in low-density medium (Balmer-dominated),
consisting of 5 SNRs only. Obviously, this sample is too small for any firm conclusions to be made. The $\Sigma-D$ slope $\beta=3.5$ obtained
 for this sample is close to that for SNRs in a dense environment and
these two classes of SNRs approximately lie on two parallel tracks or domains in the log-log plane, one
above another, as expected. However, the small number of objects in the Balmer-dominated sample strongly constrain the reliability of this relation,
therefore it should be used with caution. More optical and X-ray observations are needed for
discovering new Balmer-dominated (Ia) and oxygen-rich (Ib) SNRs.

We used our new empirical relation to estimate distances
to 147 shell-like remnants with unknown distances and obtained a drastically changed the distance scale for Galactic SNRs. Though we are aware
of theoretical and statistical flaws in the $\Sigma-D$ relation, we think that in cases where direct distance
estimates are unavailable, the $\Sigma-D$ relation should remain an important tool for distance determination.
The obtained $\Sigma-D$ relations should be used with caution, because uncertainties of distance estimates could be as large as about 50\%.
Although we have increased the number of calibrators significantly, more observations are needed to get a better understanding
of the radio evolution of SNRs.

\acknowledgments
The authors thank
Denis Leahy and Dragana Momi\'c for reading and editing the manuscript.
We thank the anonymous referee for useful suggestions which
improved quality of this paper.
This work is part of the Project No. 176005 "Emission nebulae: structure and evolution"
supported by the Ministry of Education and Science of the Republic of Serbia.
BA and BV also acknowledge financial support
through the Projects No. 176004 "Stellar
physics" and No. 176021 "Visible and invisible matter in nearby galaxies:
theory and observations", respectively.

\clearpage

\clearpage

\clearpage

\clearpage

\clearpage

\clearpage

\clearpage

\clearpage

\clearpage

\begin{deluxetable}{@{\extracolsep{0mm}}llcccc@{}}
\tablecolumns{6}

\tablecaption{Shell SNRs with known distances$^{*}$}
\vskip -2mm
\tablehead{
\colhead{} & \colhead{} & \colhead{Surface brightness} & \colhead{Distance} & \colhead{Diameter} & \colhead{} \\
\colhead{Catalog name} & \colhead{Other name} & \colhead{(Wm$^{-2}$Hz$^{-1}$sr$^{-1}$)} & \colhead{(kpc)} & \colhead{(pc)} & \colhead{Reference}
}

\startdata

G4.5+6.8$^{a}$ &  Kepler, SN1604, 3C358   &  3.18e-19	&	6.0  	&   5.2  &  1     \\
G11.2$-$0.3   &       &  1.93e-19   &   4.4   &   5.1   &    7  \\
G18.8+0.3$^{a}$ & Kes 67 & 2.66e-20   & 12.0   &    47.7     &    2    \\
G21.8$-$0.6 & Kes 69 & 2.60e-20    & 5.2  &   30.3     &     3      \\
G23.3$-$0.3 & W41 &  1.45e-20   & 4.2         &    33.0      &     2    \\
G27.4+0.4 & Kes 73, 4C$-$04.71 &  5.64e-20  &   8.65    &   10.1    &     2    \\
G31.9+0.0$^{a}$ & 3C391  &   1.03e-19   &   8.5  &    14.6     &     2  \\
G33.6+0.1$^{a}$  &  Kes 79  & 3.31e-20  & 7.0   & 20.4   &     4    \\
G41.1$-$0.3  & 3C397  &  2.94e-19  &   10.3  &  10.0  &    5    \\
G43.3$-$0.2$^{a}$ & W49B  & 4.77e-19 & 10.0 &  10.1   &    2   \\
G46.8$-$0.3$^{a}$ & HC30  & 9.53e-21  & 7.8   &   33.7  &    2    \\
G53.6$-$2.2$^{a}$ & 3C400.2, NRAO 611 & 1.30e-21 &  2.8  &  24.8   &    2 \\
G54.4$-$0.3$^{a}$  & HC40  &  2.63e-21  &   3.3   &   38.4   &    6, 7    \\
G55.0+0.3 &          &  2.51e-22  &   14.0  &   70.5  &     2  \\
G65.1+0.6  &       & 1.84e-22   &   9.0   &  175.6   &    2   \\
G74.0$-$8.5$^{a}$  &  Cygnus Loop  &  8.59e-22  &    0.54     &    30.1    &   8   \\
G78.2+2.1$^{a}$  &   $\gamma$ Cygni, DR4 & 1.34e-20  &   1.20   & 20.9   &   9 \\
G84.2$-$0.8$^{a}$  &          &   5.17e-21  & 4.50  &  23.4     &    7, 10   \\
G89.0+4.7$^{a}$  & HB21  &  3.07e-21  & 0.8  &  24.2   &    2    \\
G93.3+6.9  &  DA 530, 4C(T)55.38.1  &   2.51e-21    &    2.2  &  14.9  &    7  \\
G93.7$-$0.2  &  CTB 104A, DA 551  & 1.53e-21  & 1.5  &  34.9  &   2 \\
G94.0+1.0   &  3C434.1    &  2.61e-21  &   5.2  &  41.4   &   2 \\
G96.0+2.0  &        &   6.68e-23  &  4.0   &   30.3   &    2   \\
G108.2$-$0.6  &      &  3.19e-22  &  3.2  & 57.2  &   2  \\
G109.1$-$1.0  & CTB 109 &  4.22e-21  &  3.2  &  26.1  &    11  \\
G111.7$-$2.1$^{a}$  & Cassiopeia A, 3C461 &  1.64e-17 & 3.4  & 4.9  & 2 \\
G114.3+0.3  &       &  1.67e-22 &  0.7  &   14.3  &  2  \\
G116.5+1.1$^{a}$  &       &   3.14e-22  &  1.6  &  32.2   &  2  \\
G116.9+0.2$^{a}$  &  CTB 1  &   1.04e-21  &  1.6  &  15.8  &  2  \\
G119.5+10.2$^{a}$  & CTA 1 &  6.69e-22  &  1.4  &  36.7  &  2   \\
G120.1+1.4$^{a}$  & Tycho, 3C10, SN1572  & 1.32e-19  &  4.0   &  9.3  &  12   \\
G127.1+0.5  &  R5  &  8.92e-22 &  1.25  & 16.4  &   2    \\
G132.7+1.3$^{a}$  & HB3  & 1.06e-21  &  2.2  &  51.2  &  2   \\
G156.2+5.7$^{a}$  &      &  6.22e-23  & 1.0  &      32.0  &     13   \\
G160.9+2.6$^{a}$  &  HB9 &  9.85e-22  &   0.8   &     30.2       &     14    \\
G166.0+4.3$^{a}$  & VRO 42.05.01 &  5.47e-22   &  4.5   &  57.4   &   2   \\
G180.0$-$1.7 & S147  &   3.02e-22 &   0.62  &  32.5   &   2    \\
G189.1+3.0  &  IC443, 3C157 &  1.22E-20  &  1.5  &  19.6  &   7  \\
G205.5+0.5$^{a}$  & Monoceros Nebula &  4.98e-22 &   1.2   &   76.8   &    2 \\
G260.4$-$3.4$^{a}$ & Puppis A, MSH 08-44  & 6.52e-21 &  2.2   &   35.1  &  2 \\
G290.1$-$0.8 & MSH 11-61A & 2.38e-20 &  7.0  &   33.2   &   2    \\
G292.2$-$0.5 &        &  3.51e-21 &  8.4   &   42.3 &    2   \\
G296.5+10.0$^{a}$  &   PKS 1209-51/52 &  1.23e-21  &  2.1  &    46.7  &  15  \\
G296.8$-$0.3  & 1156-62  &  4.84e-21 &  9.6   &   46.7    &    2   \\
G309.8+0.0$^{a}$ &       &   5.39e-21  &   3.6    &   22.8  &    7  \\
G315.4$-$2.3$^{a}$ &  RCW 86, MSH 14-63  &  4.18e-21  & 2.3   &   28.1 &  2  \\
G327.4+0.4  &  Kes 27  &    1.02e-20      &    4.85  &    29.6   &    2   \\
G327.6+14.6$^{a}$  &  SN1006, PKS 1459-41  & 3.18e-21 &   2.2  &  19.2  &  2  \\
G332.4$-$0.4$^{a}$  & RCW 103  &  4.21e-20  &   3.1    & 9.0   &  2 \\
G337.0$-$0.1  & CTB 33 &  1.34e-19   &  11.0  &  4.2   &   2   \\
G337.8$-$0.1  &  Kes 41  &   5.02e-20  &   11.0  &     23.5  &   2  \\
G338.3$-$0.0  &          &   1.57e-20  &   11.0  &     25.6  &   20    \\
G340.6+0.3  &            & 2.09e-20  &    15.0  &   26.2   &     2   \\
G344.7$-$0.1  &          &  3.76e-21 &    6.3   &  18.3  &   19   \\
G346.6$-$0.2   &          &  1.88E-20   &   11.0  & 25.6 &   16    \\
G348.5+0.1$^{a}$  &  CTB 37A  &   4.82e-20  &  7.9   &    34.5    &  21   \\
G348.5+0.1$^{a}$  &  CTB 37B  &   1.35e-20  &  13.2   &    65.3   &   21   \\
G349.7+0.2$^{a}$  &           &   6.02e-19  &  18.4   &  12.0  &    2  \\
G352.7$-$0.1  &        &    1.25e-20  &  7.5  &  15.1  &    4   \\
G359.1$-$0.5$^{a}$  &        &   3.66e-21   &  7.6  &  53.1  &    17, 18  \\

\enddata
\vskip 0mm

\tablenotetext{\ }{$^{*}$ Direct distance estimates, which can be inferred from proper motions, shock and radial velocities,
HI absorption and polarization, association or interaction with HI, HII and CO molecular
clouds, OB associations, pulsars,  X-ray observations, optical extinction and low frequency radio absorption.}
\tablenotetext{\ }{ $^a$ SNRs also belonging to Case \& Bhattacharya's (1998) calibration sample}
\tablerefs{(1) Chiotellis et al. 2012; (2) Green 2009; (3) Xin et al. 2009; (4) Giacani et al. 2009; (5) Jiang et al. 2010;
(6) Junkes et al. 1992; (7) Case \& Bhattacharya 1998;  (8) Blair \& Sankrit 2005; (9) Uchiyama et al 2002; (10) Feldt \& Green 1993;
 (11) Kothes \& Foster 2012; (12) Asami Hayato et al. 2010; (13) Xu et al. 2007; (14) Leahy \& Tian 2007; (15) Giacani et al. 2000;
  (16) Hewitt et al. 2009; (17) Uchida et al. 1992b; (18) Uchida et al. 1992a; (19) Giacani et al. 2011; (20) Castelletti et al. 2011;
  (21) Tian \& Leahy 2012;}

\end{deluxetable}

\clearpage

\begin{deluxetable}{lccccc}
\tablecolumns{6}
\small
\tablewidth{0pt}
\tablecaption{The $\Sigma-D$ fit parameters for calibration sample of 60 Galactic SNRs and different regression methods.
We give $\beta$ and $A$ in relation $\Sigma_{\nu}=AD^{-\beta}$ as well as their errors.}
\tablehead{ \colhead{} & \colhead{} & \colhead{}  & \colhead{}  & \colhead{} & \colhead{Average} \\
\colhead{Fit} &  \colhead{$\beta$} & \colhead{$\Delta\beta$}  & \colhead{$\log A$ } & \colhead{$\Delta\log A$} & \colhead{fractional error} }
\startdata
OLS$(Y|X)$ & 2.2840  &  0.3282	&	-17.0028 	&   0.4895  &  0.824  \\
OLS$(X|Y)$   &  5.0375     &  0.7465   &   -13.1820 &   1.0162  & 0.469   \\
Orthogonal &  4.8161 &  0.7218 &  -13.4877    &    0.9819  & 0.472  \\
Bisector & 3.1741  & 0.3335   & -15.7641  &   0.4746  & 0.575  \\
Arithmetic mean & 3.7079 &  0.4282   & -15.0209         &    0.5817  & 0.522   \\
Geometric mean & 3.4178 &  0.3536  &   -15.4256    &  0.4898  & 0.547  \\
\enddata
\end{deluxetable}

\clearpage

\begin{deluxetable}{llccc}
\tablecolumns{5}
\small
\tablewidth{0pt}
\tablecaption{Distances to shell SNRs calculated from our $\Sigma-D$ relation}
\vskip -2mm
\tablehead{
\colhead{} & \colhead{} & \colhead{Flux density} & \colhead{Diameter} & \colhead{Distance}  \\
\colhead{Catalog name} & \colhead{Other name} & \colhead{(Jy)}  & \colhead{(pc)} & \colhead{(kpc)}
}

\startdata

G0.0+0.0    &     Sgr A East
    &     100.0       &       8.0       &       9.3   \\[0.5ex]
G0.3+0.0    &
    &     22.0       &       18.4       &       5.8   \\[0.5ex]
G1.0$-$0.1    &
    &     15.0       &       17.5       &       7.5   \\[0.5ex]
G1.4$-$0.1    &
    &     2.0       &       28.8       &       9.9   \\[0.5ex]
G1.9+0.3    &
    &     0.6       &       17.1       &        ? $^{*}$     \\[0.5ex]
G3.7$-$0.2    &
    &     2.3       &       30.5       &       8.5   \\[0.5ex]
G3.8+0.3    &
    &     3.0       &       33.6       &       6.4   \\[0.5ex]
G4.2$-$3.5    &
    &     3.2       &       39.7       &       4.9   \\[0.5ex]
G4.5+6.8    &      Kepler, SN1604, 3C358
    &     19.0       &       11.2       &       12.9 (6.0)$^{a}$ \\[0.5ex]
G4.8+6.2    &
    &     3.0       &       33.6       &       6.4   \\[0.5ex]
G5.2$-$2.6    &
    &     2.6       &       34.6       &       6.6   \\[0.5ex]
G5.5+0.3    &
    &     5.5       &       26.4       &       6.8   \\[0.5ex]
G5.9+3.1    &
    &     3.3       &       34.4       &       5.9   \\[0.5ex]
G6.1+0.5    &
    &     4.5       &       28.5       &       6.7   \\[0.5ex]
G6.4+4.0    &
    &     1.3       &       49.6       &       5.5   \\[0.5ex]
G6.5$-$0.4    &
    &     27.0       &       21.6       &       4.1   \\[0.5ex]
G7.0$-$0.1    &
    &     2.5       &       32.4       &       7.4   \\[0.5ex]
G7.2+0.2    &
    &     2.8       &       28.9       &       8.3   \\[0.5ex]
G7.7$-$3.7    &      1814$-$24
    &     11.0       &       28.1       &       4.4   \\[0.5ex]
G8.3$-$0.0    &
    &     1.2       &       23.1       &       17.7   \\[0.5ex]
G8.7$-$5.0    &
    &     4.4       &       36.1       &       4.8   \\[0.5ex]
G8.7$-$0.1    &       W30
    &     80.0       &       25.1       &       1.9   \\[0.5ex]
G8.9+0.4    &
    &     9.0       &       30.3       &       4.3   \\[0.5ex]
G9.7$-$0.0    &
    &     3.7       &       28.1       &       7.5   \\[0.5ex]
G9.8+0.6    &
    &     3.9       &       27.1       &       7.8   \\[0.5ex]
G9.9$-$0.8    &
    &     6.7       &       24.3       &       7.0   \\[0.5ex]
G10.5$-$0.0    &
    &     0.9       &       27.5       &       15.8   \\[0.5ex]
G11.0$-$0.0    &
    &     1.3       &       31.3       &       10.8   \\[0.5ex]
G11.1$-$1.0    &
    &     5.8       &       27.1       &       6.3   \\[0.5ex]
G11.1$-$0.7    &
    &     1.0       &       31.4       &       12.3   \\[0.5ex]
G11.1+0.1    &
    &     2.3       &       29.0       &       9.1   \\[0.5ex]
G11.2$-$0.3    &
    &     22.0       &       12.3       &       10.5 (4.4)$^{a}$  \\[0.5ex]
G11.4$-$0.1    &
    &     6.0       &       21.1       &       9.1   \\[0.5ex]
G11.8$-$0.2    &
    &     0.7       &       24.6       &       21.1   \\[0.5ex]
G12.0$-$0.1    &
    &     3.5       &       22.3       &       10.9   \\[0.5ex]
G12.2+0.3    &
    &     0.8       &       27.2       &       17.0   \\[0.5ex]
G12.7$-$0.0    &
    &     0.8       &       28.2       &       16.1   \\[0.5ex]
G13.5+0.2    &
    &     3.5       &       18.6       &       14.3   \\[0.5ex]
G14.1$-$0.1    &
    &     0.5       &       29.9       &       18.7   \\[0.5ex]
G14.3+0.1    &
    &     0.6       &       26.5       &       20.4   \\[0.5ex]
G15.1$-$1.6    &
    &     5.5       &       35.0       &       4.5   \\[0.5ex]
G15.4+0.1    &
    &     5.6       &       27.2       &       6.4   \\[0.5ex]
G15.9+0.2    &
    &     5.0       &       19.4       &       11.3   \\[0.5ex]
G16.0$-$0.5    &
    &     2.7       &       29.4       &       8.3   \\[0.5ex]
G16.2$-$2.7    &
    &     2.0       &       35.7       &       7.2   \\[0.5ex]
G16.4$-$0.5    &
    &     4.6       &       27.0       &       7.2   \\[0.5ex]
G17.0$-$0.0    &
    &     0.5       &       28.8       &       19.8   \\[0.5ex]
G17.4$-$2.3    &
    &     4.8       &       34.3       &       4.9   \\[0.5ex]
G17.4$-$0.1    &
    &     0.4       &       32.4       &       18.6   \\[0.5ex]
G17.8$-$2.6    &
    &     4.0       &       35.6       &       5.1   \\[0.5ex]
G18.1$-$0.1    &
    &     4.6       &       22.2       &       9.6   \\[0.5ex]
G18.6$-$0.2    &
    &     1.4       &       25.2       &       14.4   \\[0.5ex]
G18.8+0.3    &      Kes 67
    &     33.0       &       18.6       &       4.7 (12.0)$^{a}$  \\[0.5ex]
G19.1+0.2    &
    &     10.0       &       31.1       &       4.0   \\[0.5ex]
G20.4+0.1    &
    &     3.1       &       24.1       &       10.3   \\[0.5ex]
G21.0$-$0.4    &
    &     1.1       &       29.6       &       12.8   \\[0.5ex]
G21.5$-$0.1    &
    &     0.4       &       30.1       &       20.7   \\[0.5ex]
G21.8$-$0.6    &      Kes 69
    &     69.0       &       18.6       &       3.2 (5.2)$^{a}$  \\[0.5ex]
G22.7$-$0.2    &
    &     33.0       &       24.0       &       3.2   \\[0.5ex]
G23.3$-$0.3    &      W41
    &     70.0       &       21.0       &       2.7 (4.2)$^{a}$  \\[0.5ex]
G24.7$-$0.6    &
    &     8.0       &       25.6       &       5.9   \\[0.5ex]
G27.4+0.0    &    Kes 73, 4C$-$04.71
    &     6.0       &       15.9       &       13.7 (8.65)$^{a}$  \\[0.5ex]
G28.6$-$0.1    &
    &     3.0       &       27.4       &       8.7   \\[0.5ex]
G29.6+0.1    &
    &     1.5       &       23.1       &       15.9   \\[0.5ex]
G30.7+1.0    &
    &     6.0       &       31.0       &       5.1   \\[0.5ex]
G31.5$-$0.6    &
    &     2.0       &       36.5       &       7.0   \\[0.5ex]
G31.9+0.0    &      3C391
    &     24.0       &       14.1       &       8.2 (8.5)$^{a}$  \\[0.5ex]
G32.0$-$4.9    &      3C396.1
    &     22.0       &       36.6       &       2.1   \\[0.5ex]
G32.4+0.1    &
    &     0.2       &       35.6       &       20.4   \\[0.5ex]
G32.8$-$0.1    &      Kes 78
    &     11.0       &       25.3       &       5.1   \\[0.5ex]
G33.2$-$0.6    &
    &     3.5       &       32.6       &       6.2   \\[0.5ex]
G33.6+0.1    &      Kes 79, 4C00.70, HC13
    &     22.0       &       17.7       &       6.1 (7.0)$^{a}$  \\[0.5ex]
G36.6+2.6    &
    &     0.7       &       41.7       &       9.7   \\[0.5ex]
G40.5$-$0.5    &
    &     11.0       &       28.1       &       4.4   \\[0.5ex]
G41.1$-$0.3    &      3C397
    &     22.0       &       11.4       &       11.7 (10.3)$^{a}$  \\[0.5ex]
G42.8+0.6    &
    &     3.0       &       37.8       &       5.4   \\[0.5ex]
G43.3$-$0.2    &      W49B
    &     38.0       &       10.4       &       10.3 (10.0)$^{a}$  \\[0.5ex]
G43.9+1.6    &
    &     8.6       &       44.2       &       2.5   \\[0.5ex]
G45.7$-$0.4    &
    &     4.2       &       34.1       &       5.3   \\[0.5ex]
G46.8$-$0.3    &       HC30
    &     14.0       &       22.8       &       5.3 (7.8)$^{a}$  \\[0.5ex]
G49.2$-$0.7    &       W51
    &     160.0       &       18.5       &       2.1   \\[0.5ex]
G53.6$-$2.2    &      3C400.2, NRAO 611
    &     8.0       &       34.1       &       3.9 (2.8)$^{a}$  \\[0.5ex]
G54.4$-$0.3    &       HC40
    &     28.0       &       29.6       &       2.5 (3.3)$^{a}$  \\[0.5ex]
G55.0+0.3    &
    &     0.5       &       47.5       &       9.4 (14.0)$^{a}$  \\[0.5ex]
G55.7+3.4    &
    &     1.4       &       43.3       &       6.5   \\[0.5ex]
G57.2+0.8    &       4C21.53
    &     1.8       &       31.6       &       9.1   \\[0.5ex]
G59.5+0.1    &
    &     3.0       &       31.2       &       7.2   \\[0.5ex]
G65.1+0.6    &
    &     5.5       &       50.6       &       2.6 (9.0)$^{a}$  \\[0.5ex]
G65.3+5.7    &
    &     52.0       &       56.6       &       0.7   \\[0.5ex]
G67.7+1.8    &
    &     1.0       &       37.3       &       9.5   \\[0.5ex]
G69.7+1.0    &
    &     2.0       &       33.9       &       7.8   \\[0.5ex]
G73.9+0.9    &
    &     9.0       &       31.7       &       4.0   \\[0.5ex]
G74.0$-$8.5    &      Cygnus Loop
    &     210.0       &       37.1       &       0.7 (0.54)$^{a}$  \\[0.5ex]
G78.2+2.1    &      DR4, gamma Cygni SNR
    &     320.0       &       21.3       &       1.2  (1.2)$^{a}$ \\[0.5ex]
G82.2+5.3    &      W63
    &     120.0       &       29.0       &       1.3   \\[0.5ex]
G83.0$-$0.3    &
    &     1.0       &       30.2       &       13.1   \\[0.5ex]
G84.2$-$0.8    &
    &     11.0       &       25.8       &       5.0 (4.5)$^{a}$  \\[0.5ex]
G89.0+4.7    &      HB21
    &     220.0       &       28.7       &       0.9 (0.8)$^{a}$  \\[0.5ex]
G93.3+6.9 &  DA 530, 4C(T)55.38.1
    &      9.0      &       30.3    &      4.5 (2.2)$^{a}$        \\[0.5ex]
G93.7$-$0.2    &      CTB 104A, DA 551
    &     65.0       &       33.0       &       1.4 (1.5)$^{a}$  \\[0.5ex]
G94.0+1.0    &      3C434.1
    &     13.0       &       29.6       &       3.7 (5.2)$^{a}$  \\[0.5ex]
G96.0+2.0    &
    &     0.3       &       62.0       &       8.2 (4.0)$^{a}$  \\[0.5ex]
G108.2$-$0.6    &
    &     8.0       &       45.3       &       2.5 (3.2)$^{a}$  \\[0.5ex]
G109.1$-$1.0    &      CTB 109
    &     22.0       &       26.9       &       3.3 (3.2)$^{a}$  \\[0.5ex]
G111.7$-$2.1    &      Cassiopeia A, 3C461
    &     2720.0       &       5.1       &       3.5 (3.4)$^{a}$  \\[0.5ex]
G114.3+0.3    &
    &     5.5       &       51.6       &       2.5 (0.7)$^{a}$  \\[0.5ex]
G116.5+1.1    &
    &     10.0       &       45.4       &       2.3 (1.6)$^{a}$  \\[0.5ex]
G116.9+0.2    &      CTB 1
    &     8.0       &       35.7       &       3.6 (1.6)$^{a}$  \\[0.5ex]
G119.5+10.2    &      CTA 1
    &     36.0       &       39.0       &       1.5 (1.4)$^{a}$  \\[0.5ex]
G120.1+1.4    &      Tycho, 3C10, SN1572
    &     56.0       &       13.4       &       5.8 (4.0)$^{a}$  \\[0.5ex]
G126.2+1.6    &
    &     6.0       &       50.6       &       2.5   \\[0.5ex]
G127.1+0.5    &      R5
    &     12.0       &       36.8       &       2.8 (1.25)$^{a}$  \\[0.5ex]
G132.7+1.3    &      HB3
    &     45.0       &       35.5       &       1.5 (2.2)$^{a}$  \\[0.5ex]
G156.2+5.7    &
    &     5.0       &       62.9       &       2.0 (1.0)$^{a}$  \\[0.5ex]
G160.9+2.6    &      HB9
    &     110.0       &       36.1       &       1.0 (0.8)$^{a}$  \\[0.5ex]
G166.0+4.3    &      VRO 42.05.01
    &     7.0       &       40.6       &       3.2 (4.5)$^{a}$  \\[0.5ex]
G179.0+2.6    &
    &     7.0       &       49.0       &       2.4   \\[0.5ex]
G180.0$-$1.7    &      S147
    &     65.0       &       45.8       &       0.9 (0.62)$^{a}$  \\[0.5ex]
G182.4+4.3    &
    &     1.2       &       61.1       &       4.2   \\[0.5ex]
G189.1+3.0    &      IC443, 3C157
    &     160.0       &       21.8       &       1.7 (1.5)$^{a}$  \\[0.5ex]
G192.8$-$1.1    &      PKS 0607+17
    &     20.0       &       41.4       &       1.8   \\[0.5ex]
G205.5+0.5    &      Monoceros Nebula
    &     160.0       &       41.4       &       0.6 (1.2)$^{a}$  \\[0.5ex]
G206.9+2.3    &      PKS 0646+06
    &     6.0       &       43.8       &       3.1   \\[0.5ex]
G260.4$-$3.4    &      Puppis A, MSH 08$-$44
    &     130.0       &       24.6       &       1.5 (2.2)$^{a}$  \\[0.5ex]
G261.9+5.5    &
    &     10.0       &       34.3       &       3.4   \\[0.5ex]
G266.2$-$1.2    &      RX J0852.0$-$4622
    &     50.0       &       41.0       &       1.2   \\[0.5ex]
G272.2$-$3.2    &
    &     0.4       &       46.9       &       10.7   \\[0.5ex]
G279.0+1.1    &
    &     30.0       &       41.3       &       1.5   \\[0.5ex]
G284.3$-$1.8    &      MSH 10$-$53
    &     11.0       &       29.1       &       4.2   \\[0.5ex]
G286.5$-$1.2    &
    &     1.4       &       33.8       &       9.3   \\[0.5ex]
G289.7$-$0.3    &
    &     6.2       &       27.6       &       6.0   \\[0.5ex]
G290.1$-$0.8    &      MSH 11$-$61A
    &     42.0       &       19.0       &       4.0 (7.0)$^{a}$  \\[0.5ex]
G292.2$-$0.5    &
    &     7.0       &       27.9       &       5.5 (8.4)$^{a}$  \\[0.5ex]
G296.1$-$0.5    &
    &     8.0       &       34.1       &       3.9   \\[0.5ex]
G296.5+10.0    &      PKS 1209$-$51/52
    &     48.0       &       34.4       &       1.5 (2.1)$^{a}$  \\[0.5ex]
G296.8$-$0.3    &      1156$-$62
    &     9.0       &       26.2       &       5.4 (9.6)$^{a}$  \\[0.5ex]
G298.6$-$0.0    &
    &     5.0       &       24.3       &       8.0   \\[0.5ex]
G299.2$-$2.9    &
    &     0.5       &       43.7       &       10.7   \\[0.5ex]
G299.6$-$0.5    &
    &     1.0       &       36.8       &       9.7   \\[0.5ex]
G301.4$-$1.0    &
    &     2.1       &       43.9       &       5.2   \\[0.5ex]
G302.3+0.7    &
    &     5.0       &       29.6       &       6.0   \\[0.5ex]
G304.6+0.1    &      Kes 17
    &     14.0       &       17.8       &       7.6   \\[0.5ex]
G308.1$-$0.7    &
    &     1.2       &       35.5       &       9.4   \\[0.5ex]
G309.2$-$0.6    &
    &     7.0       &       25.2       &       6.4   \\[0.5ex]
G309.8+0.0    &
    &     17.0       &       25.6       &       4.0 (3.6)$^{a}$  \\[0.5ex]
G310.6$-$0.3    &      Kes 20B
    &     5.0       &       21.9       &       9.4   \\[0.5ex]
G310.8$-$0.4    &      Kes 20A
    &     6.0       &       24.8       &       7.1   \\[0.5ex]
G311.5$-$0.3    &
    &     3.0       &       20.1       &       13.8   \\[0.5ex]
G312.4$-$0.4    &
    &     45.0       &       26.3       &       2.4   \\[0.5ex]
G312.5$-$3.0    &
    &     3.5       &       33.3       &       6.0   \\[0.5ex]
G315.4$-$2.3    &      RCW 86, MSH 14$-$63
    &     49.0       &       26.9       &       2.2 (2.3)$^{a}$  \\[0.5ex]
G315.9$-$0.0    &
    &     0.8       &       44.6       &       8.2   \\[0.5ex]
G316.3$-$0.0    &       MSH 14$-$57
    &     20.0       &       24.0       &       4.1   \\[0.5ex]
G317.3$-$0.2    &
    &     4.7       &       25.2       &       7.9   \\[0.5ex]
G321.9$-$0.3    &
    &     13.0       &       29.3       &       3.8   \\[0.5ex]
G323.5+0.1    &
    &     3.0       &       29.5       &       7.8   \\[0.5ex]
G327.2$-$0.1    &
    &     0.4       &       30.1       &       20.7   \\[0.5ex]
G327.4+0.4    &      Kes 27
    &     30.0       &       22.5       &       3.7 (4.85)$^{a}$  \\[0.5ex]
G327.4+1.0    &
    &     1.9       &       33.3       &       8.2   \\[0.5ex]
G327.6+14.6    &      SN1006, PKS 1459$-$41
    &     19.0       &       28.5       &       3.3 (2.2)$^{a}$  \\[0.5ex]
G330.0+15.0    &      Lupus Loop
    &     350.0       &       32.6       &       0.6   \\[0.5ex]
G330.2+1.0    &
    &     5.0       &       24.9       &       7.8   \\[0.5ex]
G332.0+0.2    &
    &     8.0       &       23.4       &       6.7   \\[0.5ex]
G332.4$-$0.4    &      RCW 103
    &     28.0       &       16.9       &       5.8 (3.1)$^{a}$  \\[0.5ex]
G332.4+0.1    &      MSH 16$-$51, Kes 32
    &     26.0       &       20.2       &       4.6   \\[0.5ex]
G332.5$-$5.6    &
    &     2.0       &       47.7       &       4.7   \\[0.5ex]
G335.2+0.1    &
    &     16.0       &       25.5       &       4.2   \\[0.5ex]
G336.7+0.5    &
    &     6.0       &       24.7       &       7.2   \\[0.5ex]
G337.0$-$0.1    &       CTB 33
    &     1.5       &       14.2       &       32.5 (11.0)$^{a}$  \\[0.5ex]
G337.2$-$0.7    &
    &     1.5       &       24.8       &       14.2   \\[0.5ex]
G337.3+1.0    &      Kes 40
    &     16.0       &       21.3       &       5.5   \\[0.5ex]
G337.8$-$0.1    &      Kes 41
    &     18.0       &       16.3       &       7.6 (11.0)$^{a}$  \\[0.5ex]
G338.1+0.4    &
    &     4.0       &       29.5       &       6.8   \\[0.5ex]
G338.3$-$0.0    &
    &     7.0       &       20.4       &       8.8 (11.0)$^{a}$  \\[0.5ex]
G338.5+0.1    &
    &     12.0       &       19.2       &       7.3   \\[0.5ex]
G340.4+0.4    &
    &     5.0       &       22.3       &       9.1   \\[0.5ex]
G340.6+0.3    &
    &     5.0       &       19.5       &       11.2 (15.0)$^{a}$  \\[0.5ex]
G341.9$-$0.3    &
    &     2.5       &       23.8       &       11.7   \\[0.5ex]
G342.0$-$0.2    &
    &     3.5       &       26.1       &       8.6   \\[0.5ex]
G342.1+0.9    &
    &     0.5       &       37.3       &       13.5   \\[0.5ex]
G343.1$-$0.7    &
    &     7.8       &       31.0       &       4.5   \\[0.5ex]
G344.7$-$0.1    &
    &     2.5       &       27.5       &       9.5 (6.3)$^{a}$  \\[0.5ex]
G345.7$-$0.2    &
    &     0.6       &       29.9       &       17.1   \\[0.5ex]
G346.6$-$0.2    &
    &     8.0       &       19.9       &       8.5 (11.0)$^{a}$  \\[0.5ex]
G348.5$-$0.0    &
    &     10.0       &       20.8       &       7.2   \\[0.5ex]
G348.5+0.1    &      CTB 37A
    &     72.0       &       16.5       &       3.8 (7.9)$^{a}$  \\[0.5ex]
G348.7+0.3    &      CTB 37B
    &     26.0       &       21.3       &       4.3 (13.2)$^{a}$  \\[0.5ex]
G349.2$-$0.1    &
    &     1.4       &       27.3       &       12.8   \\[0.5ex]
G349.7+0.2    &
    &     20.0       &       9.9       &       15.2 (18.4)$^{a}$  \\[0.5ex]
G350.0$-$2.0    &
    &     26.0       &       31.5       &       2.4   \\[0.5ex]
G351.7+0.8    &
    &     10.0       &       25.1       &       5.4   \\[0.5ex]
G351.9$-$0.9    &
    &     1.8       &       29.9       &       9.9   \\[0.5ex]
G352.7$-$0.1    &
    &     4.0       &       21.6       &       10.7 (7.5)$^{a}$  \\[0.5ex]
G353.6$-$0.7    &
    &     2.5       &       42.9       &       4.9   \\[0.5ex]
G353.9$-$2.0    &
    &     1.0       &       36.8       &       9.7   \\[0.5ex]
G354.8$-$0.8    &
    &     2.8       &       34.8       &       6.3   \\[0.5ex]
G355.4+0.7    &
    &     5.0       &       34.6       &       4.8   \\[0.5ex]
G355.6$-$0.0    &
    &     3.0       &       22.9       &       11.3   \\[0.5ex]
G355.9$-$2.5    &
    &     8.0       &       24.2       &       6.4   \\[0.5ex]
G356.2+4.5    &
    &     4.0       &       36.2       &       5.0   \\[0.5ex]
G356.3$-$0.3    &
    &     3.0       &       25.2       &       9.9   \\[0.5ex]
G356.3$-$1.5    &
    &     3.0       &       33.1       &       6.6   \\[0.5ex]
G357.7+0.3    &
    &     10.0       &       29.6       &       4.2   \\[0.5ex]
G358.0+3.8    &
    &     1.5       &       52.3       &       4.7   \\[0.5ex]
G358.1+0.1    &
    &     2.0       &       38.1       &       6.5   \\[0.5ex]
G358.5$-$0.9    &
    &     4.0       &       31.0       &       6.3   \\[0.5ex]
G359.0$-$0.9    &
    &     23.0       &       24.6       &       3.7   \\[0.5ex]
G359.1$-$0.5    &
 &     14.0       &       27.7       &       4.0 (7.6)$^{a}$  \\[0.5ex]

\enddata
\vskip 0mm

\tablenotetext{\ }{ $^{*}$ Using our $\Sigma-D$ relation for this SNR makes no sense, since
it is reliably established that G1.9+0.3 is increasing in its flux density.}
\tablenotetext{\ }{$^a$ SNRs belonging to our new calibration sample from Table 1. Values in brackets represent directly obtained distances.}

\end{deluxetable}

\clearpage

\begin{deluxetable}{@{\extracolsep{0mm}}llcccc@{}}
\tabletypesize{\scriptsize}
\tablewidth{-20mm}
\tablecolumns{5}
\tablecaption{Basic properties of the 28 Galactic SNRs evolving in dense ISM i.e. associated with molecular clouds}
\vskip -2mm
\tablehead{
\colhead{} & \colhead{} & \colhead{} & \colhead{Surface brightness}  & \colhead{Diameter} \\
\colhead{Catalog name} & \colhead{Conventional name} & \colhead{Evidence} & \colhead{(Wm$^{-2}$Hz$^{-1}$sr$^{-1}$)} & \colhead{(pc)}
}

\startdata

G18.8+0.3$^{a}$ & Kes 67  &  CO MA \& LB, CO ratio  & 2.66e-20   &    47.7   \\
G21.8$-$0.6 & Kes 69 & 	OH, CO MA \& LB, HCO$+$, H$_{2}$  &  2.60e-20    &   30.3      \\
G23.3$-$0.3 & W41 & HI MA \& CO RC, extended TeV &  1.45e-20   &    33.0       \\
G31.9+0.0$^{a}$ & 3C391 & 	OH, molecular MA \& LB (CO, HCO$+$,CS), H$_{2}$, NIR  &   1.03e-19   &     14.6     \\
G33.6+0.1$^{a}$  &  Kes 79 & CO MA, HCO$+$ MA, broad OH absorption & 3.31e-20 & 20.4   \\
G41.1$-$0.3  & 3C397 &  CO MA \& LB &  2.94e-19 &  10.0   \\
G43.3$-$0.2$^{a}$ & W49B & 	H$_{2}$ MA, CO ratio & 4.77e-19  &  10.1    \\
G54.4$-$0.3$^{a}$  & HC40 & CO MA \& LB, IR MA &  2.63e-21   &   38.4      \\
G78.2+2.1$^{a}$  &   $\gamma$ Cygni, DR4 & 	CO MA  & 1.34e-20   & 20.9  \\
G84.2$-$0.8$^{a}$  &          & CO MA &   5.17e-21  &  23.4      \\
G89.0+4.7$^{a}$  & HB21 & CO MA \& LB, CO ratio, H$_{2}$, NIR  &  3.07e-21   &  24.2    \\
G94.0+1.0   &  3C434.1  & CO RC &  2.61e-21    &  41.4  \\
G109.1$-$1.0  & CTB 109 & CO MA \& LB  &  4.22e-21   &  26.1  \\
G111.7$-$2.1$^{a}$  & Cassiopeia A, 3C461 &  H$_{2}$CO absorption, IR RC, CO RC  &  1.64e-17   & 4.9  \\
G132.7+1.3$^{a}$  & HB3 & CO MA & 1.06e-21   &  51.2    \\
G166.0+4.3$^{a}$  & VRO 42.05.01 & unusual shape, CO RC   &  5.47e-22    &  57.4    \\
G189.1+3.0  &  IC443, 3C157 & OH, CO ratio, H$_{2}$, molecular MA \& LB  &  1.22E-20    &  19.6   \\
G205.5+0.5$^{a}$  & Monoceros Nebula &  CO RC &  4.98e-22  &   76.8    \\
G260.4$-$3.4$^{a}$ & Puppis A, MSH 08-44 &  CO RC, OH(negative) & 6.52e-21   &   35.1   \\
G290.1$-$0.8 & MSH 11-61A & CO RC  & 2.38e-20 &   33.2    \\
G332.4$-$0.4$^{a}$  & RCW 103 & 	IR MA \& colors, NIR, H$_{2}$ \& HCO$+$ MA   &  4.21e-20      & 9.0   \\
G337.0$-$0.1  & CTB 33 &  OH  &  1.34e-19   &  4.2     \\
G337.8$-$0.1  &  Kes 41 &  OH  &   5.02e-20  &     23.5  \\
G344.7$-$0.1  &          & IR RC \& colors &  3.76e-21   &  18.3   \\
G346.6$-$0.2   &          & OH, H$_{2}$, IR colors &  1.88E-20  & 25.6   \\
G348.5+0.1$^{a}$  &  CTB 37A & OH, H$_{2}$, IR MA &   4.82e-20    &    34.5 \\
G349.7+0.2$^{a}$  &           & OH, CO MA \& LB, CO ratio, H$_{2}$, IR MA  &   6.02e-19    &  12.0   \\
G359.1$-$0.5$^{a}$  &        & 	OH, CO \& H$_{2}$ MA, HCO$+$ \& CS absorption &   3.66e-21    &  53.1 \\

\enddata
\vskip 0mm

\tablenotetext{\ }{ Evidence: chief evidence that suggests the interaction between SNR and molecular clouds.
LB=line broadening, MA=morphology agreement, H$_{2}$=vibrational/rotational lines of molecular
 hydrogen [e.g. H$_{2}$ 1$-$0 S(1) line (2.12 um), H$_{2}$ 0$-$0 S(0)$-$S(7) lines], NIR=Near-Infrared
  (e.g. [Fe II] line), OH=1720 MHz OH maser, RC=rough morphological correspondence, etc. (taken from
 the List of Galactic SNRs Interacting with Molecular Clouds by Bing Jiang, \email{bjiang@nju.edu.cn}, 2010)}

\end{deluxetable}

\clearpage

\end{document}